\documentclass[aps,prb,twocolumn,superscriptaddress,footinbib]{revtex4-2}
\usepackage[utf8]{inputenc}
\usepackage[colorlinks=true,citecolor=black,urlcolor=black,linkcolor=black,pdfstartview=FitH,bookmarksopen]{hyperref}
\usepackage{graphicx}
\usepackage[nointegrals]{wasysym} 
\usepackage[export]{adjustbox}
\usepackage{amsmath,amsfonts,amssymb,latexsym}
\usepackage{hhline}
\usepackage{bm}
\usepackage{verbatim}
\usepackage{enumitem}
\usepackage{mathrsfs}
\usepackage{slashed}
\usepackage{empheq}
\usepackage{tikz}

\def\app#1#2{	\mathrel{		\setbox0=\hbox{$#1\sim$}		\setbox2=\hbox{			\rlap{\hbox{$#1\propto$}}			\lower1.1\ht0\box0		}		\raise0.25\ht2\box2	}}

\usepackage[usenames,dvipsnames]{xcolor}

\usepackage{dcolumn}

\usepackage{pifont}
\usepackage{ulem}

\usepackage{multirow}
\usepackage{braket}
\def\({\left(}
\def\){\right)}
\def\[{\left[}
\def\]{\right]}
\newcommand{\ie}{\begin{equation}\begin{aligned}}
\newcommand{\fe}{\end{aligned}\end{equation}}

\begin{document}

\title{Matrix Product States for Modulated Topological Phases:
\\
Crystalline Equivalence Principle and Lieb-Schultz-Mattis Constraints}

\author{Shang-Qiang Ning}
\email[Electronic address:$~~$]{sqning@ust.hk}
\affiliation{Department of Physics, Hong Kong University of Science and Technology, Clear Water Bay, Hong Kong, China}
\affiliation{Department of Physics, The Chinese University of Hong Kong, Sha Tin, New Territory, Hong Kong, China}

\author{Hiromi Ebisu}
\email[Electronic address:$~~$]{hiromi.ebisu@riken.jp}
\affiliation{Interdisciplinary Theoretical and Mathematical Sciences Program (iTHEMS) RIKEN, Wako
351-0198, Japan}

\author{Ho Tat Lam}
\email[Electronic address:$~~$]{hotatlam@usc.edu}
\affiliation{Department of Physics, Massachusetts Institute of Technology, Cambridge, Massachusetts 02139, USA}
\affiliation{
Leinweber Institute for Theoretical Physics, Stanford University, Stanford, CA 94305, USA}
\affiliation{Department of Physics and Astronomy, University of Southern California, Los Angeles, CA 90089, USA}

\date{\today}

\begin{abstract} 
Modulated symmetries are internal symmetries that act in a spatially non-uniform manner. Consequently, when a modulated symmetry $G_{\text{int}}$ is combined with a spatial symmetry $G_{\text{sp}}$, the total symmetry group takes the form of a semidirect product $G=G_{\text{int}}\rtimes G_{\text{sp}}$.
Using matrix product states, we classify topological phases protected by modulated symmetries together with spatial symmetries in one spatial dimension. We show that these modulated symmetry-protected topological (SPT) phases are classified by $H^{2}(G,U(1)_s)$, in agreement with the crystalline equivalence principle, which states that SPT phases protected by symmetries involving spatial elements are in one-to-one correspondence with internal SPT phases protected by the same symmetries, viewed as acting internally. Furthermore, we provide a matrix product state derivation of the Lyndon–Hochschild–Serre spectral
sequence for the corresponding internal SPT phases, which enables us to construct an explicit correspondence between modulated SPT phases and internal SPT phases. As applications of this classification, we prove a Lieb-Schultz-Mattis (LSM) theorem for modulated symmetries that forbids the existence of symmetric short-ranged entangled ground state, as well as an SPT-LSM constraint that enforces nontrivial entanglement in the SPT ground state. Finally, we use the classification to establish a similar LSM-type constraints for non-invertible Kramers-Wannier reflection symmetries.

\end{abstract}

\maketitle

\tableofcontents

\section{Introduction}
Recent developments in generalized symmetries have revealed that global symmetries can exhibit intricate interplay with spacetime structures, leading to a variety of novel phenomena such as higher-form symmetries~\cite{nussinov2009symmetry,gaiotto2015generalized} and subsystem symmetries~\cite{Ring2002,plqt_ising_2004,Vijay,seiberg2021exotic,Burnell:2021reh}. Among these, modulated symmetries—internal symmetries that transform nontrivially under spatial symmetries—have recently emerged as a natural framework for describing systems with spatially varying symmetry actions~\cite{griffin2015scalar,Pretko:2018jbi,PhysRevX.9.031035,PhysRevB.106.045112,Pace:2024tgk}. Such symmetries appear in a wide range of physical contexts, including Hilbert space fragmentation~\cite{PhysRevX.10.011047,Moudgalya_2022}, unconventional hydrodynamics~\cite{PhysRevResearch.2.033124,PhysRevResearch.3.043186}, and gauge theories with restricted mobility~\cite{Pretko_2017,Pretko_2018,seiberg2021exotic,Oh_2022,PhysRevB.106.045145,Delfino_2023,2024multipole}. Moreover, modulated symmetries naturally arise in systems with Lieb–Schultz–Mattis (LSM)–type anomalies~\cite{Lieb1961,PhysRevB.69.104431,PhysRevLett.84.1535,Cheng:2015kce,cheng2023lieb}, where gauging a combination of internal and spatial symmetries leads to symmetry structures with intrinsic spatial modulation~\cite{Aksoy:2023hve,Pace:2025hpb,ebisu2025modulated}.

Despite their growing importance, a systematic understanding of symmetry-protected topological (SPT) phases protected by modulated symmetries remains incomplete. For internal symmetries, one-dimensional bosonic SPT phases are classified by group cohomology $H^2(G_{\mathrm{int}}, U(1)_s)$~\cite{spt2013,chen2014symmetry}. There has been recent progress in approaching modulated SPT phases using matrix product states (MPS)~\cite{lam2024classification,PhysRevB.109.125121}, anomaly inflow~\cite{anomaly_2024,lam2024topological}, symmetry topological field theory~\cite{Pace:2025hpb,yao2025latticetranslationmodulatedsymmetries} and defect network constructions~\cite{bulmash2025defect}. 
 While there have been several attempts, it is not a priori clear how such classifications generalize when internal and spatial symmetries are intertwined via a nontrivial group extension.

In this work, we address this question by considering symmetry groups of the form $G = G_{\text{int}} \rtimes G_{\text{sp}}$, where an internal symmetry $G_{\text{int}}$ is acted upon nontrivially by a spatial symmetry $G_{\text{sp}}$. We develop a systematic framework to characterize one-dimensional bosonic SPT phases protected by such modulated symmetries using matrix product states (MPS). Our main result is that these modulated SPT phases are classified by the second group cohomology $H^2(G, U(1)_s)$, demonstrating that the crystalline equivalence principle continues to hold even in the presence of nontrivial mixing between internal and spatial symmetries. The crystalline equivalence principle~\cite{PhysRevX.8.011040} states that SPT phases protected by a spatial symmetry $G_{\text{sp}}$ are in one-to-one correspondence with SPT phases protected by the same symmetry group, viewed as an internal symmetry $\tilde G_{\text{sp}}$, where the orientation-reversing elements of $G_{\text{sp}}$ are mapped to anti-unitary symmetries in $\tilde G_{\text{sp}}$. When specialized to modulated symmetries, this principle predicts that the classification of modulated SPT phases is given by $H^2(\tilde G, U(1)_s)$, where $\tilde G = G_{\text{int}} \rtimes G_{\text{sp}}$, in agreement with the MPS classification, where $s:\tilde G\rightarrow \mathbb{Z}_2$ encodes which elements are anti-unitary. Notably, the MPS-based analysis naturally leads to a structure analogous to the Lyndon–Hochschild–Serre (LHS) spectral sequence for $H^2(\tilde G, U(1)_s)$, which we derive independently within the MPS framework for the corresponding internal SPT phases. This, in turn, allows us to construct an explicit correspondence between the modulated SPT phases and the internal SPT phases.

As applications of this classification, we prove a LSM theorem for modulated symmetries, which states that certain projective representations of the local symmetry operators are intrinsically incompatible with modulated SPT phases, i.e., symmetric short-range entangled ground states, and therefore imply that the symmetry must either be spontaneously broken or the system must be gapless. Interestingly, unlike the case of unmodulated symmetries, not every local projective representation leads to an LSM theorem, as previously pointed out in Ref.~\cite{pace2026lieb}. In such cases, using MPS, we show that the local projective representation instead gives rise to an SPT-LSM constraint that enforces nontrivial entanglement in the ground-state wave function. Finally, we use the classification to establish an LSM-type constraint for a class of non-invertible Kramers-Wannier reflection symmetries associated with exponential symmetries~\cite{Pace:2024tgk}, which states that the system must either spontaneously break the symmetry or be gapless.
\par
The rest of this work is organized as follows. In Sec.~\ref{sec:modulated_sym}, we review modulated symmetries and formulate their action on one-dimensional lattice systems. In Sec.~\ref{sec:MPS}, we investigate these symmetries within the MPS framework and derive the classification of modulated SPT phases. In Sec.~\ref{sec:CEP}, we provide an MPS-based derivation of the LHS spectral sequence for internal SPT phases protected by semidirect product symmetries, and use it to match with the MPS classification of modulated SPT phases, thereby establishing the crystalline equivalence principle in this context.
In Sec.~\ref{sec:ex}, we present several lattice model examples of modulated SPT phases to demonstrate the validity of our classification.
Sec.~\ref{sec:lsm} is devoted to discussing LSM anomalies and SPT-LSM constraints of modulated symmetries, as well as anomalies for a class of non-invertible Krammers-Wannier reflection symmetries. 
 Finally, in Sec.~\ref{sec:con}, we conclude our work with a few remarks. Technical details are provided in Appendix~\ref{sec:LHS}. 

\section{Modulated Symmetry}
\label{sec:modulated_sym}
Symmetries can be broadly divided into two classes: internal symmetries and spatial symmetries, denoted by $G_{\text{int}}$ and $G_{\text{sp}}$, respectively. Internal symmetries are symmetries that act on the internal degrees of freedom and do not change their spatial coordinates, while spatial symmetries are those that change the spatial coordinates, e.g.~translation or reflection symmetries. In general, these two symmetries can intertwine with each other and form a total symmetry group $G$. Since composing two internal symmetries always generates an internal symmetry, internal symmetries form a subgroup of the total symmetry group $G$. In contrast, composing two spatial symmetries may or may not generate an internal symmetry. As a result, the most general symmetry group is given by the group extension
\ie
1\rightarrow G_{\text{int}} \rightarrow G \rightarrow G_{\text{sp}} \rightarrow 1\,.
\fe

When the group extension splits, the total symmetry group $G=G_{\text{int}}\rtimes G_{\text{sp}}$ becomes a semidirect product of internal and spatial symmetry, which is specified by the action of the spatial symmetry on the internal symmetry 
\ie
\rho:\ G_{\text{sp}}\rightarrow \text{Aut}(G_{\text{int}})\,.
\fe
If this action is non-trivial, the internal symmetry necessarily acts non-uniformly in space and hence are called \textit{modulated symmetries} \cite{Pace:2024tgk}. 

We will focus on modulated symmetries on a one-dimensional lattice. We will discuss two types of spatial symmetries:~the $\mathbb{Z}$ lattice translation and the $\mathbb{Z}_2^R$ reflection. We pick the convention that the translation generator $T$ sends site $n+1$ to site $n$, while the reflection generator $R$ reflects around site $0$.

If $G_{\text{sp}}=\mathbb{Z}$ consists of only translations, the action of the spatial symmetry $G_{\text{sp}}$ on the internal symmetry $G_{\text{int}}$ is specified by the action of the translation generator
\ie
T:\ a\in G_{\text{int}}\mapsto t(a)\in G_{\text{int}}\,.
\fe
Suppose the internal symmetry operator acts in an on-site manner that takes the form 
\ie
\mathcal{U}(a)=\prod_{n} U_{n}^{(n)}(a)\,,
\fe
where $a\in G_{\text{int}}$. The subscript $n$ denotes which lattice site the local symmetry operator acts, while the superscript $(n)$ indicates that the operator may take different forms on different site.
Applying the translation generator on the symmetry operator gives
\ie\label{eq:translation_generator}
T\, \mathcal{U}(a) T^{-1}=\prod_{n} U_{n}^{(n+1)}(a)\,,
\fe
which should agree with
\ie
\mathcal{U}({t(a)})
= \prod_{n} U_{n}^{(n)}(t(a))\,.
\fe
Comparing the two expressions, we find the relation
\ie
U^{(n+1)}(a)=U^{(n)}(t(a))\,.
\fe
Applying this relation recursively, we can express the symmetry operator in terms of $U^{(0)}(a)$ and the translation map $T$ as
\ie\label{eq:sym_op}
\mathcal{U}(a)=\prod_{n} U_{n}^{(0)}(t^n(a))\,.
\fe

If $G_{\text{sp}}=\mathbb{Z}_2^R$ consists of only reflection, the action of the spatial symmetry $G_{\text{sp}}$ on the internal symmetry $G_{\text{int}}$ is specified by the action of the reflection generator
\ie\label{eq:reflection_generator}
R:\ a\in G_{\text{int}}\mapsto r(a)\in G_{\text{int}}\,.
\fe
Applying the reflection generator on the symmetry operator gives
\ie
R\, \mathcal{U}(a) R= \prod_{n} U^{(n)}_{-n}(a)\,,
\fe
which should agree with
\ie
\mathcal{U}(r(a))=\prod_{n} U^{(n)}_{n}(r( a))\,.
\fe
Comparing the two expressions, we find the relation
\ie\label{eq:reflection_constraint}
U^{(n)}(r( a))=U^{(-n)}(a)\,.
\fe
At $n=0$, it becomes a constraint on $U^{(0)}(a)$
\ie\label{eq:reflection_constraint_site_0}
U^{(0)}(a)=U^{(0)}(r( a))\,.
\fe

We emphasize that the modulation of the internal symmetry is meaningful only in the presence of spatial symmetries as previously discussed in \cite{bulmash2025defect}. As an example, consider the case when the internal symmetry group $G_{\text{int}}$ is finite. In this case, the translation homomorphism must have a finite order. Therefore, there exists an integer $N$ such that $t^N(a)=a$, which implies that the symmetry operator $\mathcal{U}(a)$ has a finite periodicity of $N$ and thus can be interpreted as an unmodulated symmetry after blocking $N$ lattice sites into one. Such symmetries are called $N$-cycle symmetries in \cite{Sauerwein:2019ugp,Stephen:2019zas,Stephen:2022zyt}. However, such blocking breaks the lattice translation symmetry down to the subgroup generated by $T^N$. Thus, one cannot distinguish a modulated and unmodulated symmetry unless the spatial symmetry is present.

\section{MPS for Modulated SPT Phases}\label{sec:MPS}
In this section, we use MPS to classify 1+1D topological phases protected by modulated symmetries together with lattice symmetry. We will refer to these SPT phases as modulated SPT phases, although as we discussed in the previous section spatial symmetry is crucial. 

MPS has been employed to classify ordinary SPT phases \cite{Chen:2010zpc,Pollmann_2010}.
The basis for using MPS to classify SPT phases is the theorem, stating that any ground state of a one-dimensional gapped local Hamiltonian can be efficiently approximated by MPS \cite{Fannes:1990ur}. An MPS is a special class of state built out of the MPS tensors $A^i_{\alpha \beta}$:
\ie
A^{i}_{\alpha\beta}=\!
\begin{tikzpicture}[scale=1.2, every node/.style={font=\small},baseline={(A.base)}]
    \node[draw, circle, minimum size=0.6cm] (A) at (0,0) {};
    \node[] at (0,0) {$A$};

    \draw (-0.6,0) -- (A.west);
    \draw (A.east) -- (0.6,0);
    \draw (A.north) -- (0,0.6);

    \node[left] at (-0.6,0) {$\alpha$};
    \node[right] at (0.6,0) {$\beta$};
    \node[above] at (0,0.6) {$i$};
\end{tikzpicture}\,.
\fe
For each physical index $i$, $A^i$ is a $D\times D$ matrix with the virtual indices $\alpha,\beta$. The MPS is built by contracting the virtual indices $\alpha,\beta,...$ of these MPS tensors as
\ie
|\Psi\rangle=\sum_i \text{Tr}[A^{i_1}_1\cdots A^{i_L}_L]\,
|i_1,\cdots,i_L\rangle\,.
\fe
Here, the MPS tensors are allowed to vary from site to site, as indicated by the subscript of the tensor.  As the ground state of modulated SPT phases is unique, the MPS can be chosen to be injective. 
An MPS is injective if its transfer matrix 
\ie\label{eq:transfer}
E_n=\sum_i A_n^{i\dagger} A_n^i
\fe
has a unique eigenvalue of maximum modulus equals to 1. Without the loss of generality, we can choose the left eigenvector to be the identity matrix and the maximum modulus eigenvalue to be 1. Such MPS are called left-canonical.
Injective MPS are particularly nice as they obey the fundamental theorem of MPS \cite{Perez-Garcia:2008hhq}, which states that two (non-translation invariant) injective MPS of the left canonical form define the same state if their MPS tensors, $A_n$ and $B_n$, are related by 
\ie\label{eq:MPS_theorem}
\begin{tikzpicture}[scale=1.2, every node/.style={font=\small},baseline={(A.base)}]
    \node[draw, circle, minimum size=0.6cm] (B) at (0,0) {};
    \node[] at (0,0) {$B_n$};

    \draw (-0.6,0) -- (B.west);
    \draw (B.east) -- (0.6,0);
    \draw (B.north) -- (0,0.6);
\end{tikzpicture}
\ = \
e^{i\theta_n}\!\begin{tikzpicture}[scale=1.2, every node/.style={font=\small},baseline={(A.base)}]
    \node[draw, circle, minimum size=0.6cm] (A) at (0,0) {};
    \node[] at (0,0) {$A_n$};

    \draw (-0.6,0) -- (A.west);
    \draw (A.east) -- (0.6,0);
    \draw (A.north) -- (0,0.6);

    \node[left] at (-0.5,0) {$V_n$};
    \node[right] at (0.5,0) {$V_{n+1}^{\dagger}$};
\end{tikzpicture}\,,
\fe
where $V_n$ are unitary operators.
This theorem will be important for our classification of modulated SPT.

\subsection{Translation}

After this brief review of MPS, we are now ready to discuss MPS for modulated SPT phases. We will first discuss the case when the lattice symmetry $G_{\text{sp}}=\mathbb{Z}$ consists of only lattice translations. We will also assume $G_{\text{int}}$ is unitary throughout.

In this case, the modulated symmetry operators $\mathcal{U}(a)$ take the form of Eq.~\eqref{eq:sym_op}, and the MPS tensor for the symmetric ground state $|\Psi\rangle$ of the modulated SPT phases
can be chosen to be translation invariant, since the ground state preserves translation symmetry. Acting these symmetry operators $\mathcal{U}(a)$ on the MPS ground state $|\Psi\rangle$ gives
\ie\label{eq:U_g_psi}
\sum_i\, 
\text{Tr}{\left[ \left(U(a)\cdot A\right)^{i_1}\cdots \left(U(t^{L}(a))\cdot A\right)^{i_L}\right]}\, |i_1,\cdots,i_L\rangle\,,
\fe
where we omitted the superscript of $U^{(0)}$ and used the short-hand notation $\left(U\cdot A\right)^{i}:=\sum_j U^{i}{}_{j}A^{j}$.
We can interpret $U(t^{n}(a))\cdot A$ as the new MPS tensor at site $n$ after the symmetry action. The ground state should be invariant under the action of $\mathcal{U}(a)$ up to a phase.
Thus, by the fundamental theorem of MPS in Eq.~\eqref{eq:MPS_theorem}, we have the following push-through equality
\ie\label{eq:push_through}
U(t^{n}(a)) \cdot A =e^{i\theta_{n}(a)}\, V_{n}(a)\, A\, V_{n+1}^{\dagger}(a)\,.
\fe
Since the MPS tensor is translation invariant, we can view the left-hand side as pushing the symmetry operator associated with $t^n(a)$ at site 0 and thus we have an alternative push-through equility
\ie
U(t^{n}(a))\cdot A=e^{i\theta_{0}(t^n(a))}\, V_{0}(t^n(a))\,A\, V_{1}^{-1}(t^n(a))\,.
\fe
Comparing the two expressions, we arrive at the following recursion relations
\ie
e^{i\theta_{n}(a)}=e^{i\theta_{0}(t^n(a))}\,,
\quad
V_{n}(a)=V_{0}(t^n(a))\,.
\fe
Substituting them back to Eq.~\eqref{eq:push_through}, we arrive at the following push-through equation
\ie
\begin{tikzpicture}[scale=1.2, every node/.style={font=\small},baseline={(A.base)}]
    \node[draw, circle, minimum size=0.6cm] (A) at (0,0) {};
    \node[] at (0,0) {$A$};

    \draw (-0.6,0) -- (A.west);
    \draw (A.east) -- (0.6,0);
    \draw (A.north) -- (0,0.6);

    \node[above] at (0,0.55) {$U(a)$};
\end{tikzpicture}=e^{i\theta_{0}(a)}\begin{tikzpicture}[scale=1.2, every node/.style={font=\small},baseline={(A.base)}]
    \node[draw, circle, minimum size=0.6cm] (A) at (0,0) {};
    \node[] at (0,0) {$A$};

    \draw (-0.6,0) -- (A.west);
    \draw (A.east) -- (0.6,0);
    \draw (A.north) -- (0,0.6);

    \node[left] at (-0.55,0) {$V_{0}(a)$};
    \node[right] at (0.55,0) {$V_{0}(t(a))^{\dagger}$};
\end{tikzpicture}\,.
\fe

Assuming that $U(a)$ acts linearly at every site, the action of $U(ab)$ on the MPS tensor should agree with that of acting $U(a)$ and $U(b)$ successively. The case where $U(a)$ acts projectively will be discussed in Sec.~\ref{sec:lsm_constraint}. Comparing the two actions gives the following equality
\ie\label{eq:MPS_consistency}
&e^{i\theta_0(ab)}\, V_0(ab)\, A\,V_0(t(ab))^\dagger=
\\
&e^{i\theta_0(a)}e^{i\theta_0(b)}\, V_0(a)V_0(b) \,A\,V_0(t(b))^\dagger V_0(t(a))^\dagger\,.
\fe
For the two expressions to agree, $V_0(a)$ can at most furnish a projective representation of $G_{\text{int}}$\,:
\ie\label{eq:projective_rep}
V_0(a)V_0(b)=\omega(a,b)V_0(ab)\,.
\fe
Associativity of the group multiplication  requires these projective phases to satisfy the 2-cocycle condition
\ie
\omega(a,bc)\omega(b,c)=\omega(a,b)\omega(ab,c)\,.
\fe
Substituting Eq.~\eqref{eq:projective_rep} into Eq.~\eqref{eq:MPS_consistency}, we obtain
\ie\label{eq:linearity}
\boxed{\frac{\omega(a,b)}{\omega(t(a),t(b))}=\frac{e^{i\theta_0(ab)}}{e^{i\theta_0(a)}e^{i\theta_0(b)}}\,. }
\fe
The combination on the right-hand side is a 2-coboundary, which implies that $\omega(a,b)$ and $\omega(t(a),t(b))$ belong to the same cohomology class. Consequently, only those classes in $H^2(G_{\text{int}},U(1))$ that are invariant under the translation map $T$ correspond to valid modulated SPT phases. We refer to such classes as the \textit{strong indices} of the modulated SPT phases. Physically, these strong indices can be detected by placing the modulated SPT phase on an open chain, where the virtual operators $V(a)$ becomes the boundary operators and the strong index is reflected in their projectivity. A non-trivial strong index implies symmetry-protected edge modes, which lead to a ground state degeneracy. The strong index can also be detected by examining the entanglement spectrum of the ground state wave function. A non-trivial strong index leads to a degeneracy in the entanglement spectrum \cite{Pollmann_2010}.

After specifying a strong index, there can still exist distinct modulated SPT phases.
These are related by multiplying $e^{i\theta_0(a)}$ by a one-dimensional representation $e^{i\alpha(a)}$ of  $G_{\text{int}}$
\ie
e^{i\theta_0(a)} \mapsto e^{i\theta_0'(a)} =e^{i\theta_0(a)} e^{i\alpha(a)}\,.
\fe
The one-dimensional representation $e^{i\alpha(a)}$ can also be interpreted as a 1-cocycle in $H^1(G_{\text{int}},U(1))$, obeying the 1-cocycle condition
\ie
\frac{e^{i\alpha(ab)}}{e^{i\alpha(a)}e^{i\alpha(b)}}=1\,.
\fe
Thus, the multiplication does not spoil the condition in Eq.~\eqref{eq:linearity}.
However, different choices of $e^{i\alpha(a)}$ do not necessarily correspond to distinct SPT phases, since we are free to redefine the virtual operator $V_0(a)$ by attaching to it a one-dimensional representation $\mu(a)$ of $G_{\text{int}}$
\ie
V_0(a)\mapsto \mu(a)V_0(a)\,.
\fe
Such a redefinition generates the equivalence relation
\ie\label{eq:equivalence_translation}
\boxed{e^{i\alpha(a)}\sim e^{i\alpha(a)}\frac{\mu(a)}{\mu(t(a))}\,.}
\fe
Therefore, distinct modulated SPT phases with the same strong index are classified by the equivalence classes in $H^1(G_{\text{int}},U(1))/(T^*-1)$, where $T^*$ denotes the pullback of $T$ on cohomology. We refer to these equivalence classes as the \textit{weak indices} of the modulated SPT phases. We emphasize, for a fixed strong index, the set of modulated SPT phases forms a torsor rather than a group. These weak indices do not affect the symmetry algebra localized on the boundary when we place the modulated SPT phase on an open chain. 

They can however be detected by examining how translation acts in the presence of a modulated symmetry defect. Let us place the modulated SPT phase on an infinite chain. We can create a symmetry defect of $a$ between site $n$ and $n+1$ by applying the symmetry operator $a$ to every site to the left of site $n$. The translation generator $T$ translates the symmetry defect by one site. To restore the location of the defect, we accompany $T$ with a local symmetry operator $U_n(t^{n+1}(a))$. The combined translation operator 
\ie
\tilde T=U_n(t^{n+1}(a))\,T
\fe
therefore keeps the defect at the same location, while transforming its type from $a$ to $t(a)$. Acting with this modified translation operator $\tilde T$ on the ground state of the defect Hilbert space then produces a phase $e^{i\theta_0(t^n(a))}$, which directly measures the weak index.

In summary, modulated SPT phases are classified by
\ie
\boxed{H^2(G_{\text{int}},U(1))^T\times \frac{H^1(G_{\text{int}},U(1))}{T^*-1}\,,}
\fe
where $H^2(G_{\text{int}},U(1))^T$ denotes the collection of second cohomology classes that are invariant under the $T^*$ action
\ie
H^2(G_{\text{int}},U(1))^T=\left\{[\omega]\in H^2(G_{\text{int}},U(1))\,\vert\, T^*[\omega]=[\omega]\right\}\,.\label{h2}
\fe
Our classification agrees with the previous result obtained using the other methods \cite{bulmash2025defect,yao2025latticetranslationmodulatedsymmetries}.
In Sec.~\ref{sec:CEP}, we will show that this classification is identical to the classification from crystalline equivalence principle, which  gives $H^2(G,U(1))=H^2(G_{\text{int}}\rtimes\mathbb{Z},U(1))$.

\subsection{Reflection}
We now turn to the case when the lattice symmetry $G_{\text{sp}}=\mathbb{Z}_2^R$ consists of only reflection. In this case, the form of the modulated symmetry operator is not fixed to the one in Eq.~\eqref{eq:sym_op}. But the compatibility with the reflection symmetry still imposes constraint on the symmetry operators as Eqs.~\eqref{eq:reflection_constraint} and~\eqref{eq:reflection_constraint_site_0}.

In 1+1D, reflection symmetry alone already distinguishes two SPT phases \cite{Pollmann_2012,Pollmann_2010}. We briefly review the argument using the MPS tensor following \cite{Pollmann_2012,Pollmann_2010}.
Under reflection, the MPS ground state is mapped to
\ie
R|\Psi\rangle = \sum_i \text{Tr}[(A^T)^{i_1}\cdots (A^T)^{i_L}]\,
|i_1,\cdots,i_L\rangle\,.
\fe
By the fundamental theorem of MPS in Eq.~\eqref{eq:MPS_theorem}, the MPS is reflection symmetric if its MPS tensors satisfy
\ie
A^T=e^{i\theta_r}V_r A V_r^\dagger\,, \label{eq:reflection_MPS}
\fe
where $A^T$ denotes the transpose of $A$. 
Applying this equation twice, we arrive at 
\ie
A=e^{2i\theta_r}(V_r^* V_r) A (V_r^* V_r)^\dagger\,.
\fe
This equation implies that $(V_r^* V_r)^\dagger$ is a left eigenvector of the transfer matrix $E_n$ in Eq.~\eqref{eq:transfer} with eigenvalue $e^{2i\theta_r}$. However, as a left-canonical injective MPS, the identity matrix is only eigenvector of the transfer matrix with eigenvalue of modulus 1 and the eigenvalue is 1. Thus, we have $e^{2i\theta_r}=1$ and
\ie\label{eq:V_r}
V_r^* V_r= \omega(r,r)\ \Rightarrow\ V_r^T = \omega^*(r,r) V_r\,,
\fe
where $\omega(r,r)$ is a phase. Applying the second equation twice, we obtain $V_R = \omega^*(r,r) ^2 V_R$. Thus, $\omega(r,r)$ can only take two discrete values $+1$ or $-1$, which characterizes the two distinct reflection SPT phases.

Let us now discuss the constraint of reflection symmetry on the push-through property of the internal symmetry. To this end, we place the modulated SPT phase on an open chain, with sites running from $n=-L$ to $n=L$ so that the reflection symmetry around $n=0$ is preserved. Since $\mathcal{U}(r(a))R=R\,\mathcal{U}(a)$, pushing both sides through the MPS tensors should give the same virtual operators on the boundary. On the left hand side, we have
\ie
&(\mathcal{U}(r(a)) R)\cdot \mathcal{A} 
=\mathcal{U}(r(a))\cdot \mathcal{A}^T
\\
=\,&e^{i\theta_r} V(r)\,\big(\mathcal{U}(r(a))\cdot \mathcal{A}\big)\, V_r^\dagger
\\
=\,&e^{i\theta_r}e^{i\theta(r(a))} V(r) V_{-L}(r(a))\,\mathcal{A}\, V_L(r(a))^\dagger V_r^\dagger\,,
\fe
where we use $\mathcal{A}$ to denote the products of all the MPS tensors from $n=-L$ to $n=L$ and use $e^{i\theta(a)}$ to denote the product of all $e^{i\theta_n(a)}$.
On the right hand side, we have
\ie
&(R \,\mathcal{U}(a))\cdot \mathcal{A} 
=R\cdot (U(a) \cdot \mathcal{A})
\\
=\, &e^{i\theta(a)}\,R\cdot\left(V_{-L}(a) \,\mathcal{A}\, V_L(a)^\dagger\right)
\\
=\, &e^{i\theta(a)}\,V_{L}(a)^* \,\mathcal{A}^T\, V_{-L}(a)^T
\\
=\, &e^{i\theta_r}e^{i\theta(a)}\,V_{L}(a)^* V_r \, \mathcal{A}\, V_r^\dagger V_{-L}(a)^T\,.
\fe
Comparing the two expressions, we find that
\ie
e^{i\theta(r(a))}=e^{i\theta(a)}\,,\quad   V_{-L}(r(a))=V(r)^\dagger V_{L}(a)^* V_r\,.
\label{eq:theta_condition}
\fe
This means that when we push $\mathcal{U}(a)$ through the MPS tensors, we have
\ie
\begin{tikzpicture}[scale=1.2, every node/.style={font=\small},baseline={(A.base)}]
    \node[draw, circle, minimum size=0.6cm] (A) at (0,0) {};
    \node[] at (0,0) {$\mathcal{A}$};

    \draw (-0.6,0) -- (A.west);
    \draw (A.east) -- (0.6,0);
    \draw (A.north) -- (0,0.6);

    \node[above] at (0,0.55) {$\mathcal{U}(a)$};
\end{tikzpicture}=e^{i\theta(a)}\begin{tikzpicture}[scale=1.2, every node/.style={font=\small},baseline={(A.base)}]
    \node[draw, circle, minimum size=0.6cm] (A) at (0,0) {};
    \node[] at (0,0) {$\mathcal{A}$};

    \draw (-0.6,0) -- (A.west);
    \draw (A.east) -- (0.6,0);
    \draw (A.north) -- (0,0.6);

    \node[left] at (-0.55,0) {$V_r^\dagger V_{L}(r(a))^*V_r$};
    \node[right] at (0.55,0) {$V_{L}(a)^\dagger$};
\end{tikzpicture}\,.
\fe
Since $\mathcal{U}(a)$ acts linearly, applying $\mathcal{U}(ab)$ on the MPS should agree with applying $\mathcal{U}(a)$ and $\mathcal{U}(b)$ succesively. The former gives
\ie
e^{i\theta(ab)}\, V_r^\dagger V_{L}(r(ab))^*V_r\, A \, V_L(ab)^\dagger\,,
\fe
while the later gives
\ie
e^{i\theta(a)}e^{i\theta(b)}\, V_r^\dagger V_{L}(r(a))^* V_{L}(r(b))^*V_r \, A \, V_L(b)^\dagger V_L(a)^\dagger\,.
\fe
The two expressions agree if the projective phase, i.e., the 2-cocycle, appearing in the virtual operator algebra
\ie
V_L(a)V_L(b)=\omega(a,b)V_L(ab)\,,
\fe
satisfies the condition
\ie
\boxed{\omega(a,b)\,\omega(r(a),r(b))=\frac{e^{i\theta(a)}e^{i\theta(b)}}{e^{i\theta(ab)}}\,.}\label{eq:reflection_cocycle_condition}
\fe
It implies that $\omega(a,b)$ and $\omega(r(a),r(b))$ belong to inverse cohomology classes. Consequently, 
only those classes in $H^2(G_{\text{int}},U(1))$ that are mapped to their inverse under the reflection map $R$ correspond to valid modulated SPT phases.
We refer to such classes as the strong indices of the modulated SPT phases. Similar to the strong indices for the case with translation symmetry, these strong indices can be detected by the projectivity of the boundary symmetry operators and the degeneracy in the entanglement spectrum.

After specifying a strong index, there can still exist distinct modulated SPT phases related by multiplying $e^{i\theta(a)}$ by one-dimensional representation $e^{i\alpha(a)}$ of  $G_{\text{int}}$, i.e., 1-cocyle in $H^1(G_{\text{int}},U(1))$,
\ie
e^{i\theta(a)} \mapsto e^{i\theta'(a)} =e^{i\theta(a)} e^{i\alpha(a)}\,.
\fe
This modification does not spoil the condition in Eq.~\eqref{eq:reflection_cocycle_condition}. However, for it to preserve the condition in Eq.~\eqref{eq:theta_condition}, we require
\ie
\boxed{e^{i\alpha(r(a))}=e^{i\alpha(a)}\,,}
\fe
which means that the 1-cocycle has to be invariant under the reflection map $R$. However, different choices of $e^{i\alpha(a)}$ do not necessarily correspond to distinct SPT phases, since we are free to redefine the virtual operator $V_L(a)$ by attaching to it a one-dimensional representation $\mu(a)$
\ie
V_L(a)\rightarrow \mu(a)^*V_L(a)\,.
\fe
Such a redefinition generates the equivalence relation
\ie
\boxed{e^{i\alpha(a)}\sim e^{i\alpha(a)}\mu(a)\mu(r(a))\,.}\label{eq:trivialization_reflection}
\fe
Therefore, distinct modulated SPT phases with the same strong index are classified by the equivalence classes in $H^1(G_{\text{int}},U(1))^R/(R^*+1)$, where $R^*$ denotes the pullback of $R$ on cohomology and
\ie
&H^1(G_{\text{int}},U(1))^R=
\\
&\left\{e^{i\alpha(a)}\in H^1(G_{\text{int}},U(1))\,\vert\, e^{i\alpha(r(a))}=e^{i\alpha(a)}\right\}\,.
\fe
We refer to these equivalence classes as the weak indices of the modulated SPT phases since they do not affect the projectivity of the boundary symmetry operators. 

They can however be detected by examining how reflection acts in the presence of a modulated symmetry defect. As before, we place the modulated SPT phase on an open chain with sites labeled by $n=-L,...,L$. We insert a symmetry defect of type $a$ between sites $n=-1$ and $n=0$ by applying the truncated symmetry operator $\mathcal{U}(a)_{-L,-1}$ on sites $n=-L$ through $n=-1$.
The reflection generator $R$ maps $\mathcal{U}(a)_{-L,-1}$ to $\mathcal{U}(r(a))_{1,L}$, which acts on sites $n=1$ through $n=L$.
To restore the location of the defect, we need to accompany $R$ with a local symmetry operator $U_0^{(0)}(r(a))$ at the origin, and define the combined reflection operator
\ie
\tilde R=U_0^{(0)}(r(a)) \, R\,.
\fe
This combined reflection maps $\mathcal{U}(a)_{-L,-1}$ to 
\ie
\mathcal{U}(r(a))_{0,L}=\mathcal{U}(r(a))_{-L,-1}^\dagger \mathcal{U}(r(a))\,.
\fe
The truncated operator $\mathcal{U}(r(a))_{-L,-1}^\dagger$ can be interpreted as creating a defect of type $r(a)^{-1}$ between sites $n=-1$ and $n=0$. Thus, the combined reflection leaves the defect location invariant, while transforming the defect type from $a$ to $r(a)^{-1}$. The additional symmetry operator
$\mathcal{U}(r(a))$ then picks up a phase $e^{i\theta(r(a))}$ when acting on the ground state, thereby measuring the weak index.

In summary, the modulated SPT phases in the presence of reflection symmetry are classified by
\ie
\boxed{H^2(G_{\text{int}},U(1))^R\times \frac{H^1(G_{\text{int}},U(1))^R}{R^*+1}\times\mathbb{Z}_2\,,}
\fe
where $H^2(G_{\text{int}},U(1))^R$ denotes the collection of second cohomology classes that are mapped to their inverse under the $R^*$ action
\ie
H^2(G_{\text{int}},U(1))^R=\left\{[\omega]\in H^2(G_{\text{int}},U(1))\,\vert\, R^*[\omega]=[\omega^{-1}]\right\}\,.
\fe
In Sec.~\ref{sec:CEP}, we will show that this classification is identical to the classification from crystalline equivalence principle, which  gives $H^2(G,U(1))=H^2(G_{\text{int}}\rtimes\mathbb{Z}_2^R,U(1)_s)$.

\section{Matching the Crystalline Equivalence Principle}
\label{sec:CEP}

The crystalline equivalence principle~\cite{PhysRevX.8.011040} states that the classification of bosonic SPT phases protected by spatial symmetry $G_{\text{sp}}$ on a contractible spatial manifold is isomorphic to that of SPT phases protected by an internal symmetry~$\tilde{G}_{\text{sp}}$. The internal symmetry group $\tilde{G}_{\text{sp}}$ is related to the spatial symmetry group ${G}_{\text{sp}}$ by mapping reflection of the latter to time reversal of the former. 

In this section, we establish the crystalline equivalence principle for modulated SPT phases. Specifically, we construct an explicit map between modulated SPT phases and internal SPT phases protected by the symmetry $\tilde G=G_{\text{int}}\rtimes \tilde{G}_{\text{sp}}$, which are classified by $H^2(\tilde G, U(1)_s)$. Here, $s: \tilde G\rightarrow\mathbb{Z}_2$ is a homomorphism that encodes whether a symmetry element is unitary or anti-unitary, thereby defining a $G$-action on the $U(1)$ coefficient. We assume $G_{\text{int}}$ is unitary throughout, so any anti-unitary element in $\tilde G$ can only come from reflection symmetries in $G_{\text{sp}}$. For $\mathbf{g}\in \tilde G_{\text{sp}}$, we assign $s(\mathbf{g})=+1$ if $\mathbf{g}$ is unitary and $s(\mathbf{g})=-1$ if it is anti-unitary.

\subsection{MPS Derivation of 
LHS Spectral Sequence}

To start with, we use MPS to derive the classification of internal SPT phases protected by the symmetry $\tilde G=G_{\text{int}}\rtimes \tilde{G}_{\text{sp}}$, which reproduces the LHS spectral sequence for $H^2(\tilde G, U(1)_s)$. Our strategy is to first consider a $G_{\text{int}}$ SPT and then deduce what conditions are imposed on it from the $\tilde G_{\text{sp}}$ symmetry.

We will denote the group elements of $G_{\text{int}}$ by $a,b,c$, the group elements of $\tilde G_{\text{sp}}$ by $\mathbf{g}, \mathbf{h},\mathbf{k}$ and the group elements of $\tilde G$ by $a_\mathbf{g},b_\mathbf{h},c_\mathbf{k}$. The image of $a\in G_{\text{int}}$ under the action of $\mathbf{g}\in \tilde G_{\text{sp}}$ will be denoted by ${}^{\mathbf{g}}a$.

Without loss of generality, we assume the internal symmetry operators $\mathcal{U}(a_\mathbf{g})$ are translation invariant and act in an on-site manner. We denote the corresponding local symmetry operator by $U(a_\mathbf{g})$.
The MPS tensor of a $ G_{\text{int}}$ SPT ground state satisfies the push-through property 
\begin{align}
    U(a)\cdot A
    = e^{i\theta(a)} V(a)AV^\dagger(a)\,.
    \label{eq:pullthrough}
\end{align}
Although $U(a)$ form a linear representation of $G_{\text{int}}$, $V(a)$ alone can form a projective representation:
\begin{align}
     V(a) V(b)=\omega(a,b) V(ab) \,.\label{eq:phy_gen0}
 \end{align}
with the projective phase $\omega(a,b)$ satisfying the 2-cocycle condition
\begin{align}
\boxed{\,
\frac{\omega(a,b)\omega(ab,c)}{\omega(a,bc) \omega(b,c)}=1}\,.\label{eq:phy_gener_0}
\end{align}

Now, consider the action of $\tilde G_{\text{sp}}$ on the $G_{\text{int}}$ SPT ground state. For the ground state to be symmetric under $\tilde G_{\text{sp}}$, the MPS tensor must satisfy a similar push-through property as in Eq.~\eqref{eq:pullthrough} for local symmetry operator $U(\mathbf{g})$, with $V(\mathbf{g})$ the corresponding virtual operator. Note that $V(\mathbf{g})$ is anti-unitary if $\mathbf{g}$ is anti-unitary.
Following the group multiplication, we have
\ie
U(\mathbf{g}) U(a) U(\mathbf{g})^{-1}= U({}^\mathbf{g}a)\,.
\fe
Applying both sides to the MPS tensor leads to an analogous algebra for $V(a)$ and $V(\mathbf{g})$, with the freedom to include an additional phase:
\begin{align}
&V(\mathbf{g}) V(a) V(\mathbf{g})^{-1}= e^{i\phi_\mathbf{g}(a)}V({}^\mathbf{g}a) \,.\label{eq:phy_gen_def_phi}
\end{align}

Next, consider the conjugated action of $V(\mathbf{g})$ on both sides of Eq.~\eqref{eq:phy_gen0}. Using Eq.~\eqref{eq:phy_gen_def_phi}, the left hand side gives
 \ie
    & V(\mathbf{g}) V(a) V(b) V(\mathbf{g})^{-1} 
    \\
    =\,&e^{i(\phi_\mathbf{g}(a)+\phi_\mathbf{g}(b))} V({}^\mathbf{g}a)V({}^\mathbf{g}b) \\
    =\,&e^{i(\phi_\mathbf{g}(a)+\phi_\mathbf{g}(b))}
 \omega({}^\mathbf{g}a,{}^\mathbf{g}b) V({}^\mathbf{g}a{}^\mathbf{g}b)\,,
 \fe
while the right hand side gives
\ie
      &V(\mathbf{g})\omega(a,b)V(ab)V(\mathbf{g})^{-1}\\
      =\,&\omega^{s(\mathbf{g})}(a,b) e^{i\phi_\mathbf{g}(ab)} V({}^\mathbf{g}a{}^\mathbf{g}b)\,.
 \fe
The exponent $s(\mathbf{g})$ indicates that $\omega(a,b)$ is complex conjugated whenever $V(\mathbf{g})$ is anti-unitary.
 Comparing the two expressions, we obtain the condition
 \ie
     \boxed{\,\frac{\omega({}^\mathbf{g}a,{}^\mathbf{g}b)}{ \omega^{s(\mathbf{g})}(a,b)}=\frac{e^{i\phi_\mathbf{g}(ab)}}{e^{i\phi_\mathbf{g}(a)}e^{i\phi_\mathbf{g}(b)}}}\,.\label{eq:phy_gener_1}
 \fe

Secondly, consider the conjugate action of $V(\mathbf{g})$ and $V(\mathbf{h})$ on $V(a)$, which should agree with that of $V(\mathbf{g}\mathbf{h})$. Applying $V(\mathbf{g})$ and $V(\mathbf{h})$ successively  leads to 
\ie
&V(\mathbf{g}) V(\mathbf{h}) V(a) V(\mathbf{h})^{-1} V(\mathbf{g})^{-1} \\
=\, & V(\mathbf{g}) e^{i\phi_\mathbf{h}(a)}V({}^\mathbf{h}a)  V(\mathbf{g})^{-1}\\
=\, &  e^{i s(\mathbf{g})\phi_\mathbf{h}(a)} e^{i\phi_\mathbf{g}({}^\mathbf{h}a)} V({}^\mathbf{gh}a)  \,,
\fe
while applying $V(\mathbf{g}\mathbf{h})$ directly leads to
\begin{align}
V(\mathbf{g}\mathbf{h}) V(a) V(\mathbf{g}\mathbf{h})^{-1} 
= e^{i \phi_\mathbf{gh}(a)} V({}^\mathbf{gh}a)\,. 
\end{align}
Consistency between the two expressions then requires
\begin{align}
\boxed{\,\frac{e^{is(\mathbf{g})\phi_\mathbf{h}(a)}  e^{i\phi_\mathbf{g}({}^\mathbf{h}a)}}{e^{i \phi_\mathbf{gh}(a)}}=1}\,.\label{eq:phy_gener_2}
\end{align}

Lastly, there can be a projective phase in the algebra of the virtual operators $V(\mathbf{g})$,
\begin{align}
V(\mathbf{g})V(\mathbf{g})=\omega(\mathbf{g},\mathbf{h}) V(\mathbf{gh})\,,
\end{align}
which satisfies the following associativity
\begin{align}
\boxed{\frac{\omega(\mathbf{g},\mathbf{h}) \omega(\mathbf{gh},\mathbf{k})}{\omega(\mathbf{g},\mathbf{hk}) \omega^{s(\mathbf{g})}(\mathbf{h},\mathbf{k})}=1\,.}\label{eq:phy_gener_3}
\end{align}

In summary, we have derived four conditions Eqs.~\eqref{eq:phy_gener_0}, \eqref{eq:phy_gener_1}, \eqref{eq:phy_gener_2}  and \eqref{eq:phy_gener_3} for the given $G_{\text{int}}$ SPT to be lifted to a valid $\tilde G$ SPT. 
These conditions precisely match those required by the LHS spectral sequence (see Appendix~\ref{sec:LHS}), which enables the construction of a valid 2-cocycle $\omega(a_\mathbf{g},b_\mathbf{h})$ of $\tilde G$ from $\omega(a,b), \phi_\mathbf{g}(a)$ and $\omega(\mathbf{g},\mathbf{h})$.
This 2-cocycle is the projective phase appearing in the algebra of generic virtual operators $V(a_\mathbf{g})$:
\begin{align}
V(a_\mathbf{g}) V(b_{\mathbf{h}})=\omega(a_\mathbf{g},b_\mathbf{h}) V(a_\mathbf{g}\times b_\mathbf{h}) \label{eq:phy_general0}\,.
\end{align}
We can derive its explicit form by expanding a generic virtual operators in terms of those for $G_{\text{int}}$ and $\tilde G_{\text{sp}}$
\ie
V(a_\mathbf{g})=V(\mathbf{g})V({}^{\overline{\mathbf{g}}}a)\,.
\fe
Using this expansion, the left hand side of Eq.~\eqref{eq:phy_general0} becomes
\ie
&V(a_\mathbf{g})V(b_\mathbf{h})
\\
=\,&V(\mathbf{g})V({}^{\overline{\mathbf{g}}}a)V(\mathbf{h})V({}^{\overline{\mathbf{h}}}b)
\\
=\,&V(\mathbf{g})\,e^{-i\phi_\mathbf{h}({}^{\overline{\mathbf{gh}}}a)}V(\mathbf{h}) V({}^{\overline{\mathbf{gh}}}a)\,V({}^{\overline{\mathbf{h}}}b)
\\
=\, &e^{-is(\mathbf{g})\phi_\mathbf{h}({}^{\overline{\mathbf{g}}}a)}\,V(\mathbf{g})V(\mathbf{h}) \,V({}^{\overline{\mathbf{gh}}}a)V({}^{\overline{\mathbf{h}}}b)
\\
=\, &e^{-is(\mathbf{g})\phi_\mathbf{h}({}^{\overline{\mathbf{g}}}a)}\,\omega(\mathbf{g},\mathbf{h})\,\omega({}^{\overline{\mathbf{gh}}}a,{}^{\overline{\mathbf{h}}}b) \, V(\mathbf{gh}) V({}^{\overline{\mathbf{gh}}}a{}^{\overline{\mathbf{h}}}b)\,,
\,.
\fe
while the right hand side becomes
\ie
&\omega(a_\mathbf{g},b_\mathbf{h}) V(a_\mathbf{g}\times b_\mathbf{h})
\\
=\,& \omega(a_\mathbf{g},b_{\mathbf{h}})V((a\,{}^\mathbf{g}b)_{\mathbf{g}\mathbf{h}})
\\
=\,&\omega(a_\mathbf{g},b_{\mathbf{h}}) V(\mathbf{gh}) V({}^{\overline{\mathbf{gh}}}a{}^{\overline{\mathbf{h}}}b)\,.
\fe
Comparing the two expressions, we obtain
\begin{align}\label{eq:parametrization_2_cocycle}
\omega(a_\mathbf{g},b_\mathbf{h})=\omega({{}^{\overline{\mathbf{gh}}}a,^{\overline{\mathbf{h}}}b}) \times e^{-is(\mathbf{g})\phi_\mathbf{h}({}^{\overline{\mathbf{gh}}}a)}  \times \omega({\mathbf{g},\mathbf{h}})\,.
\end{align}
This expansion of $\omega(a_\mathbf{g},b_\mathbf{h})$ also matches the one appearing in the LHS spectral sequence (see Appendix~\ref{sec:LHS}).

\begin{figure}
    \centering
    \includegraphics[width=0.9\linewidth]{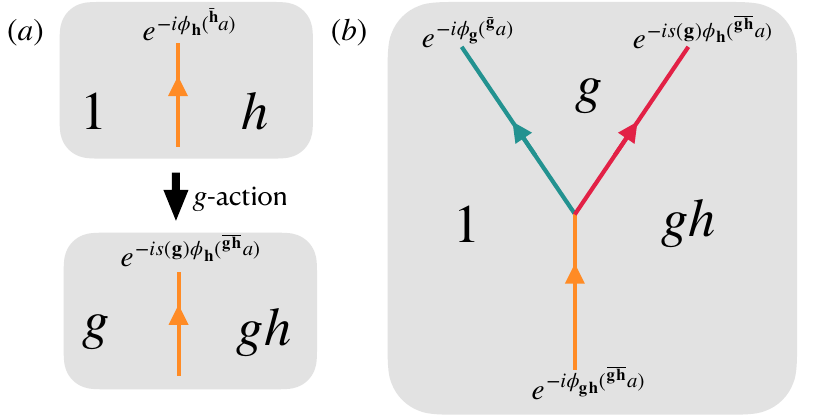}
    \caption{(a) On the oriented $\mathbf{h}$ domain wall, whose left/right side corresponds to the $\mathbf{1}$/$\mathbf{h}$ domain, we decorate with a one-dimensional representation of $G_{\text{int}}$ given by $\omega_2(a,1_\mathbf{h})=e^{-i\phi_\mathbf{h}({}^{\overline{\mathbf{h}}}a)}$.  Upon applying a $\mathbf{g}$-action, the left/right domain is mapped to the $\mathbf{g}$/$\mathbf{gh}$ domain. Meanwhile, the decorated one-dimensional representation transforms as $\mathbf{g}\cdot \omega_2(a,1_\mathbf{h})=\omega_2({}^{\overline{\mathbf{g}}}a,1_\mathbf{h})^{s(\mathbf{g})}= e^{-is(\mathbf{g})\phi_\mathbf{h}({}^{\overline{\mathbf{gh}}}a)}$. (b) This decoration rule is consistent with the fusion of domain walls, as precisely captured by Eq.~\eqref{eq:phy_gener_2}. 
    }
    \label{fig:fusion_DW}
\end{figure}

Physically, $\omega(a,b)=\omega(a_\mathbf{1},b_\mathbf{1})$ and $\omega(\mathbf{g},\mathbf{h})=\omega(1_\mathbf{g},1_\mathbf{h})$ characterize the projectivity of $G_{\text{int}}$ and $\tilde G_{\text{sp}}$ on the boundary of the $\tilde G$ SPT, respectively. Meanwhile, the phase $e^{-i\phi_\mathbf{g}({}^{\overline{\mathbf{g}}}a)}=\omega(a,1_\mathbf{g})$ admits a physical interpretation as decorating the $\mathbf{g}$ domain wall in the bulk with a 0+1D $G_{\text{int}}$ SPT, i.e., a $G_{\text{int}}$ charge, in accordance with the decorated domain wall picture underlying the LHS spectral sequence \cite{QR21}. The condition in Eq.~\eqref{eq:phy_gener_2} ensures that this domain wall decoration respects the fusion rule, as illustrated in Fig.~\ref{fig:fusion_DW}.

Importantly, not every $\phi_\mathbf{g}(a)$ corresponds to a distinct SPT. This is because we can redefine the virtual operator $V(a)$ by attaching to it a one-dimensional representation $\mu(a)$ of $G_{\text{int}}$,
\ie
V(a) \mapsto V'(a)=V(a) \mu(a)\,.
\fe
A simple physical interpretation of such redefinition is to create a pair of $G_{\text{int}}$ charge, $\mu(a)$ and $\mu(a)^{-1}$, in the bulk, and move them to the two ends. By Eq.~\eqref{eq:phy_gen_def_phi}, this induces an equivalence relation among $\phi_\mathbf{g}(a)$ as
\begin{align}
\boxed{e^{i \phi_\mathbf{g}(a)}\sim e^{i\phi'_\mathbf{g}(a)}=e^{i \phi_\mathbf{g}(a)}\frac{\mu^{s(\mathbf{g})}(a)}{\mu({}^\mathbf{g}a)}\,.}
\end{align}
In particular, $\phi_\mathbf{g}(a)$ represents a trivial $\tilde G$ SPT if it satisfies the following \textit{trivialization condition}, 
\begin{align}
e^{i \phi_\mathbf{g}(a)}=\frac{\mu({}^\mathbf{g}a)}{\mu^{s(\mathbf{g})}(a)}\,.
\label{eq:trivialize_cond}
\end{align}

\begin{figure}
    \centering
    \includegraphics[width=0.9\linewidth]{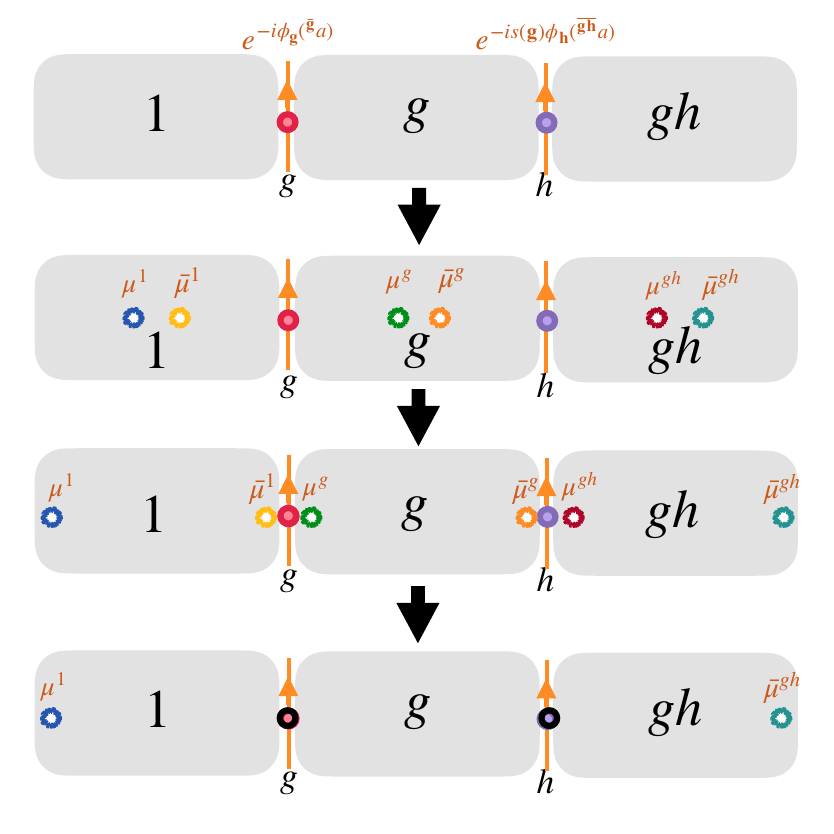}
    \caption{Trivialization of the weak modulated SPT. If the 0+1D decoration data $\omega_2(a,1_\mathbf{g})=e^{-i\phi_\mathbf{g}({}^{\overline{\mathbf{g}}}a)}$ takes the form  $\mu(a)/[\mu({}^{\overline{\mathbf{g}}}a)]^{s(\mathbf{g})}$ for some  $\mu\in H^1(G_{\text{int}}, U(1))$, then the decoration can be continuously removed while preserving the symmetry.  However, for a system with open boundary conditions, this procedure leaves behind a nontrivial 0+1D boundary state. Such a boundary state is referred to as an anomalous SPT. Here, we denote $\mu \equiv \mu^{\mathbf{1}}$.
    }
    \label{fig:Trivialization}
\end{figure}

A physical interpretation of trivialization is illustrated in Fig.~\ref{fig:Trivialization}. Let us first recall that $\omega({}^\mathbf{g}a,1_\mathbf{g})=e^{-i\phi_\mathbf{g}(a)}$ describes the decoration of 0+1D $G_{\text{int}}$ SPT on the domain wall labeled by $\mathbf{g}$, as discussed in  Ref.~\cite{QR21}.  The trivialization condition in Eq.~\eqref{eq:trivialize_cond} then implies that when the decoration takes this particular form, the corresponding domain-wall-decorated state can be continuously connected to the trivial undecorated state. This continuous process is illustrated in Fig.~\ref{fig:Trivialization}. First, we create a pair of one-dimensional representations of $G_{\text{int}}$, $\mu^\mathbf{g}$ and $\bar \mu^\mathbf{g}$, at each site with $\mathbf{g}$ state.  Then, we move this pair to the neighboring domain wall, sending $\mu^{\mathbf{g}}$ to the left and $\bar\mu^{\mathbf{g}}$ to the right. Now, at each domain wall, the total one-dimensional representation of $G_{\text{int}}$ consists of three contributions:~$\omega({a,1_\mathbf{g}})$, $\bar\mu(a)$ and $\mu^\mathbf{g}(a)=[\mu({}^{\overline{\mathbf{g}}}a)]^{s(\mathbf{g})}$. If $\omega({a,1_\mathbf{g}})$ satisfies the trivialization condition Eq.~\eqref{eq:trivialize_cond}, the combined one-dimensional representation is trivial, and hence domain wall no longer carries any decoration. 

Interestingly, when such 1+1D SPT is placed on an open chain, after removing the domain wall decoration in the bulk, there are still 0+1D boundary degrees of freedom leaving behind, which are characterized by the one-dimensional representation $\mu(a)$ of $G_{\text{int}}$.  Such 0+1D states are called anomalous SPT phases in Ref.~\cite{QR21} whose anomaly stems from the fact that they cannot exist on their own without the ``trivial" bulk. This is because the 0+1D $G_{\text{int}}$ SPT specified by $\mu(a)$ cannot be promoted to a 0+1D $\tilde G$ SPT. Mathematically, this is the statement that the one-dimensional representation $\mu(a)$ of $G_{\text{int}}$ cannot be lifted to a one-dimensional representation of the full group $\tilde G$.

\subsection{ 
Translation} 

Let us now apply the crystalline equivalence principle to the case, where the spatial symmetry $G_{\text{sp}}=\mathbb{Z}$ consists of only translations. The corresponding internal symmetry group is then $\tilde{G}_{\text{sp}}=\mathbb{Z}$, whose elements we denote by $n,m \in \mathbb{Z}$.

According to Eq.~\eqref{eq:phy_gener_2}, the phase $e^{i\phi_n(a)}$ satisfies
\begin{align}
    e^{i\phi_{n+1}(a)}=e^{i\phi_{1}(a)}e^{i\phi_{n}(t(a))} \,.\label{eq:translation_2}
\end{align}
For $n=0$, it implies $\phi_0(a)=0$. For $n=-1$, it further implies $\phi_{-1}(a)=-\phi_1(\bar t(a))$, where $\bar t$ denotes the inverse of $t$. By solving the equation recursively, we can relate $\phi_{n}$ for all $n\neq 0$ to $\phi_1$ as
\ie
    \phi_{n}(a)=
    \begin{dcases}
    \sum_{k=0}^{n-1} \phi_1(t^k(a))\ \ &\text{for }n\geq 1
    \\
    -\sum_{k=1}^{|n|} \phi_1(\bar t^{k}(a))\ &\text{for }n\leq -1
    \end{dcases}\,.\label{eq:expres}
\fe
According to Eq.~\eqref{eq:phy_gener_1}, the 2-cocycle $\omega(a,b)$ of the $G_{\text{int}}$ SPT satisfies
\begin{align}
    \frac{\omega(t^{n}(a),t^{n}(b))}{\omega({a,b})}=\frac{e^{i\phi_n(ab)}}{e^{i\phi_n(a)}e^{i\phi_n(b)}}\label{eq:LHS_2}\,,
\end{align}
which means that  $\omega({t^n}(a),{t^n}(b))$ and $\omega(a,b)$ belong to the same cohomology class in $H^2(G_{\text{int}}, U(1))$.  This condition is equivalent to the special case of $n=1$,
\begin{eqnarray}
   \frac{\omega(t (a), t(b))}{\omega(a,b)}=\frac{e^{i\phi_1(ab)}}{e^{i\phi_1(a)}e^{i\phi_1(b)}}\,,\label{strong}
\end{eqnarray}
since the condition for general $n$ follows by recursion.
The last piece of data is the 2-cocycle $\omega(n,m)$ of $\mathbb{Z}$, which can be taken to be $\omega(n,m)=1$, since $H^2(\mathbb{Z}, U(1))=\mathbb{Z}_1$. 
 Finally, the trivialization condition leads to the equivalence relation 
\begin{align}
   e^{i\phi_n(a)}\sim e^{i\phi'_n(a)}=e^{i\phi_n(a)}\frac{\mu(t^n(a))}{\mu(a)}\,,
   \label{eq:trivialize_cond_translation}
\end{align}
where $\mu(a)\in H^1(G_{\text{int}},U(1))$.

Now we establish the explicit crystalline equivalence correspondence between the modulated SPT phases and internal SPT phases. The straightforward cases are those protected purely by $G_{\text{int}}$ or purely by $G_{\text{sp}}$ (equivalently, $\tilde G_{\text{sp}}$). For SPT phases protected purely by $G_{\text{int}}$, the 2-cocycles $\omega \in H^2(G_{\text{int}}, U(1))$ satisfy the same condition, Eq.~\eqref{eq:linearity} or Eq.~\eqref{strong}, in both settings.  Therefore, the sets of allowed $\omega$ are identical and can be naturally identified. For SPT phases protected purely by translation symmetry or by an internal $\mathbb{Z}$ symmetry, only the trivial phase exists, and hence the correspondence is trivial. 

The more intricate part of the correspondence concerns the weak modulated SPT phases protected jointly by $G_{\text{int}}$ and translation symmetry, and the internal SPT phases protected jointly by $G_{\text{int}}$ and an internal $\mathbb{Z}$ symmetry. According to our classification, the former are characterized by $e^{i\theta_0(a)}\in H^1(G_{\text{int}},U(1))$, while the latter are characterized by $e^{i\phi_1(a)}\in H^1(G_{\text{int}},U(1))$. We claim that the explicit correspondence is given by
\begin{align}\label{eq:correspondence}
    \theta_0(a)\  
    \longleftrightarrow\
    \phi_1(a) \,,
\end{align}
which is consistent with Eqs.~\eqref{eq:linearity} and~\eqref{strong}. Furthermore, both data satisfy the same equivalence relation, Eq.~\eqref{eq:equivalence_translation} or Eq.~\eqref{eq:trivialize_cond_translation}.

We now provide a physical justification for this correspondence. Recall that the physical meaning of $\phi_n(a)$ is the one-dimensional representation of $G_{\text{int}}$ decorating the $\mathbb{Z}$ domain walls. For a $-n$ oriented domain wall with left/right side being the $0$/$-n$ domain, the decoration is given by $\omega(a,1_{-n})=e^{-i\phi_{-n}(t^n(a))}$, where \ie
    -\phi_{-n}(t^n(a))=
    \begin{dcases}
    \sum_{k=0}^{n-1} \phi_1(t^k(a))\ \ &\text{for }n\geq 1
    \\
    -\sum_{k=1}^{|n|} \phi_1(\bar t^{k}(a))\ &\text{for }n\leq -1
    \end{dcases}\,.\label{eq:phi_n}
\fe
On the other hand, the physical meaning of $\theta_n$ is the one-dimensional representation of $G_{\text{int}}$ located at site $n$. The ``domain wall" of translation symmetry labeled by $-n$  can be viewed as the region between site $0$  and site $n$.  Note that according to our convention, the translation generator $T$ maps site $n$ to $n-1$, so site $n$ is reached by applying $T^{-n}$ to site $0$. The total charge accumulated in this domain wall region is:
\ie
     \Theta_{-n}^{(0)}(a)=\begin{dcases}
     \sum_{k=0}^{n-1} \theta_k(a)
    = \sum_{k=0}^{n-1} \theta_0\!\left(t^{k}(a)\right)
    &\text{for } n \ge 1
    \\
    -\sum_{k=1}^{|n|} \theta_{-k}(a)
    = -\sum_{k=1}^{|n|} \theta_0\!\left(\bar t^{k}(a)\right)
    &\text{for } n \leq -1
    \end{dcases}\,, \label{eq:Theta_n}
\fe
where the minus sign for $n\leq 1$ reflects the orientation reversal of the domain wall. The superscript $(0)$ of $\Theta_{-n}^{(0)}(a)$ indicates that the domain wall is based at site $0$. Comparing Eq.~\eqref{eq:Theta_n} with Eq.~\eqref{eq:phi_n}, we find that the identification in Eq.~\eqref{eq:correspondence} ensures that the charge accumulated in the domain wall region matches the domain-wall decoration. This thus establishes the explicit correspondence between weak modulated SPT phases and the corresponding internal SPT phases.

We remark that the translation domain wall can also be located between sites $m$ and $m+n$, which can be obtained by applying the translation symmetry operator $T^{-m}$ to the domain wall between site $0$ and $n$. In that case, the charge on the domain wall becomes 
\begin{align}
     \Theta_{-n}^{(m)}(a)=\begin{dcases}
     \sum_{k=0}^{n-1} \theta_0\!\left(t^{m+k}(a)\right)
    &\text{for } n \ge 1
    \\
     -\sum_{k=1}^{|n|} \theta_0\!\left(t^{m-k}(a)\right)
    &\text{for } n \leq -1
    \end{dcases}\,.
\end{align}
Correspondingly, on the internal SPT side, the one-dimensional representation on the $-n$ domain wall is also acted upon by the internal symmetry action of $m\in\mathbb{Z}$. As a result, the domain wall charge becomes
\begin{align}
    \omega_2({t}^m(a),1_{-n})=e^{-i\phi_{-n}(t^{m+n}(a))}\,,
\end{align}
which agrees with the transformation on the modulated SPT side.

\subsection{Reflection}
Let us now turn to the case, where $G_{\text{sp}}=\mathbb{Z}_2^R$ includes only reflection. By the crystalline equivalence principle, the corresponding internal symmetry $\tilde G_{\text{sp}}=\mathbb{Z}_2^{T}$ is an anti-unitary symmetry. 

The generic condition in Eq.~\eqref{eq:phy_gener_1} implies that the 2-cocycle $\omega(a,b)$ of $G_{\text{int}}$ SPT should satisfy
 \begin{align}
     \omega(a,b)\omega(r(a),r(b))=\frac{e^{i\phi_r(ab)}}{e^{i\phi_r(a)}e^{i\phi_r(b)}}\,,\label{eq:phy_deri1}
 \end{align}
 which means that $\omega(a,b)$ and $\omega(r(a),r(b))$ belong to the inverse cohomology class of $H^2(G_{\text{int}}, U(1))$. 
 Secondly, according to the generic condition in Eq.~\eqref{eq:phy_gener_2},  the phase $\phi_r(a)$ should satisfy
 \begin{align}
    e^{i\phi_r(a)}=e^{i\phi_r(r(a))}\,,
    \label{eq:phy_deri2}
\end{align} 
which means that $e^{i\phi_r(a)}$ is invariant under the reflection action.
Trivialization induces an equivalence relation among these phases
\begin{align}
e^{i\phi_r(a)}\sim e^{i\phi_r(a)}\mu(a)\mu(r(a))\,,\label{eq:trivialization_phi(r(a))}
\end{align}
where $\mu(a)\in H^1(G_{\text{int}}, U(1))$.
In addition to these data, there is an additional $\mathbb{Z}_2$ invariant arising from 
$H^2(\mathbb{Z}_2^{T},U(1)_s)=\mathbb{Z}_2$, corresponding to a nontrivial time-reversal SPT phase, characterized by $\omega(r,r)=-1$. 

We now establish an explicit correspondence between modulated SPT phases protected by $G_{\text{int}}\rtimes \mathbb{Z}_2^R$ and internal SPT phases protected by $G_{\text{int}}\rtimes \mathbb{Z}_2^T$. First, the projective representation $\omega(a,b)$ that characterizes the $G_{\text{int}}$ SPT phases in both cases, satisfies exactly the same algebraic condition, Eq.~\eqref{eq:reflection_cocycle_condition} or Eq.~\eqref{eq:phy_deri1}. So their classification data are identical, which leads to a natural correspondence. Second, the SPT phases protected purely by time reversal symmetry form a $\mathbb{Z}_2$ classification, while those protected purely by reflection symmetry likewise form a $\mathbb{Z}_2$ classification. The correspondence between these two $\mathbb{Z}_2$ factors is therefore straightforward:~the nontrivial time reversal SPT is mapped to the nontrivial reflection SPT. Both of them are characterized by $\omega(r,r)=-1$. 
Finally, the correspondence between the modulated SPT phases protected jointly by $G_{\text{int}}$ and $\mathbb{Z}_2^R$ and the internal SPT phases protected jointly by $G_{\text{int}}$ and $\mathbb{Z}_2^T$ can be directly established by mapping $\theta(a)$, defined in Eq.~\eqref{eq:reflection_MPS}, with $-\phi_r(a)$, defined in Eq.~\eqref{eq:phy_deri1}. These two data satisfy the same algebraic relations, Eqs.~\eqref{eq:theta_condition} and \eqref{eq:reflection_cocycle_condition} or Eqs.~\eqref{eq:phy_deri1} and \eqref{eq:phy_deri2}, as well as the equivalence relations, Eq.~\eqref{eq:trivialization_reflection} or Eq.~\eqref{eq:trivialization_phi(r(a))}.

\section{Examples}\label{sec:ex}
To illustrate how the general classification derived in the previous sections is realized in concrete systems, we study explicit lattice models exhibiting modulated symmetries. In particular, we consider two representative classes of such symmetries:~\textit{exponential symmetry} and \textit{dipole symmetry}. In the following, we analyze their SPT classifications and provide microscopic realizations. 
\subsection{Exponential SPT Phases}
We first consider the case of exponential symmetry. The internal symmetry is $G_{\text{int}}=\mathbb{Z}_N\times\mathbb{Z}_N$, with group elements denoted by
$g=(g_1,g_2)$.
The lattice translation acts nontrivially on the internal symmetry as 
\begin{eqnarray}
    (g_1,g_2)\mapsto (ag_1,bg_2)\label{a,b}\,,
\end{eqnarray}
where $a$ and $b$ are positive integers coprime to $N$. For a finite periodic system of length $L$, 
consistency requires
\begin{eqnarray}
    a^L=b^L=1\pmod N\,.\label{L}
\end{eqnarray}
It implies that $L$ must be a multiple of $\text{lcm}(r,s)$, where lcm stands for the least common multiple and
\begin{eqnarray*}
    r=\text{ord}_N(a),\quad 
     s=\text{ord}_N(b)\label{r,s}
\end{eqnarray*}
denote the multiplicative orders of $a$ and $b$ in $\mathbb{Z}_N$. They are defined as $\text{ord}_N(a)=\text{min}\{k\,|\, a^k=1 \pmod N\}$ and similarly for $\text{ord}_N(b)$. This type of
 modulated symmetry is referred to as \textit{exponential symmetry} since the symmetry operator depends exponentially on the lattice site. 

\subsubsection{Classification}
To evaluate the strong index of the SPT phases protected by the exponential symmetry, consider the group cohomology of the internal symmetry, $H^2(G_{\mathrm{int}}, U(1))$, which is generated by the $2$-cocycle
\begin{eqnarray}
\omega(g,h) = \exp\left(\frac{2\pi i}{N}  kg_1 h_2\right)\,.
\label{g,h}
\end{eqnarray}
Here, $g=(g_1,g_2)$ and $h=(h_1,h_2)$ are elements of $G_{\mathrm{int}}=\mathbb{Z}_N\times\mathbb{Z}_N$, and $k \in \mathbb{Z}_N$. For the cocycle to define a valid modulated SPT phase, it must be invariant under the translation action, i.e., $\omega(g,h) = \omega(t(g), t(h))$. Using the translation action in Eq.~\eqref{a,b}, this condition becomes
\begin{eqnarray}
\exp\left(\frac{2\pi i }{N} k g_1 h_2\right)
=
\exp\left(\frac{2\pi i }{N} ab k g_1 h_2\right)\,,
\end{eqnarray}
which implies $k(ab-1)=0\pmod N$. Hence, the strong index is given by $\mathbb{Z}_{(ab-1,N)}$, where $(i,j)$ denotes the greatest common devisor of $i$ and $j$.
\par
Weak SPT indices correspond to one-dimensional representations of $G_{\text{int}}$. These are labeled by
\begin{eqnarray}
    \chi_{s,t}(g)=\exp\left({\frac{2\pi i}{N}(sg_1+tg_2)}\right)\,,
\end{eqnarray}
with $(s,t)\in\mathbb{Z}_N\times\mathbb{Z}_N$. However, 
as we discussed around Eq.~\eqref{eq:equivalence_translation}, 
two such representations may correspond to the same modulated SPT phase if they are related by translation. 
The translation action in Eq.~\eqref{a,b} thus induces the following equivalence relation on the weak indices:
\begin{eqnarray*}
    (s,t)\sim (as,bt)\, \Longrightarrow\, ((a-1)s,(b-1)t)\sim(0,0)\,.\label{condition}
\end{eqnarray*}
Quotienting by this relation, the weak index is given by $\mathbb{Z}_{(a-1,N)}\times\mathbb{Z}_{(b-1,N)}$. 
\par
Overall, combining the strong and weak indices yields
\ie
 &H^2(G_{\text{int}}\rtimes\mathbb{Z},U(1))=\\
&\mathbb{Z}_{(ab-1,N)}\times \mathbb{Z}_{(a-1,N)}\times\mathbb{Z}_{(b-1,N)}\,.\label{80}
\fe
This result agrees with the classification derived from the LHS spectral sequence in the previous section. 

\subsubsection{Lattice Models}\label{sec:exp_lattice}
We now construct an explicit lattice realization of these SPT phases. Consider a 1+1D lattice with periodic boundary condition, where each site hosts two $\mathbb{Z}_N$ qudits. Let 
$X_n, Z_n$ and $\tilde{X}_n, \tilde{Z}_n$ denote the generalized Pauli operators acting on the two qudits at site $n$, respectively. 
The exponential symmetry is realized as
\begin{eqnarray}
    \mathcal{U}_a=\prod_{n} X_n^{a^n}\,,\quad     \mathcal{U}_b=\prod_{n} \tilde{X}_n^{b^n}\,,\label{symmetry}
\end{eqnarray}
which transforms under translation as in Eq.~\eqref{a,b} 
\ie
T \,\mathcal{U}_a T^{-1}&= (\mathcal{U}_a)^a\,,
\quad
T \,\mathcal{U}_b T^{-1}&= (\mathcal{U}_b)^b\,.\label{eq:exp_algebra_translation}
\fe

Following the decorated domain wall construction~\cite{chen2014symmetry} and its generalization to modulated SPT phases~\cite{Han:2023fas}, the Hamiltonian for strong exponential SPT phases can be constructed by decorating the charge operator $\tilde X_n$ of $\,\mathcal{U}_b$ on the domain wall operator $(Z_{n-1}^{ a}Z_n^{\dagger })^p$ of $\,\mathcal{U}_a$, and vice versa:
\ie
    H=-\sum_{n} \left[Z_{n-1}^{p a}\tilde{X}_nZ_n^{\dagger p}+\tilde{Z}_n^{ qb}X_n\tilde{Z}_{n+1}^{\dagger  q}\right]+\text{h.c.}\label{exspt}
\fe
where $p,q\in\mathbb{Z}_N$. For the Hamiltonian to describe an SPT phase, we require all the terms in the Hamiltonian to commute so that the ground state is unique and gapped. This imposes the constraints
\ie
p+bq&=0\pmod N\,,
\\
pa+q&=0\pmod N\,.
\fe
Th solutions take the form
\ie
q=\frac{k N}{(ab-1,N)}\,,\quad p=-\frac{k bN}{(ab-1,N)}\,,
\fe
labeled by $k\in \mathbb{Z}_{(ab-1,N)}$. The distinct solutions correspond to distinct strong SPT phases in agreement with the classification we obtained in Eq.~\eqref{80}.

Since all terms in the Hamiltonian commute, the ground state is the unique state satisfying, for all $n$,
\begin{eqnarray}
    Z_{n-1}^{pa}\tilde{X}_nZ_n^{\dagger p}\ket{\text{GS}}
=\tilde{Z}_n^{ qb}X_n\tilde{Z}_{n+1}^{\dagger q}\ket{\text{GS}}=\ket{\text{GS}}.\label{gs}
\end{eqnarray}
It can be constructed explicitly by noting that there exists a finite-depth circuit mapping the SPT Hamiltonian to a trivial paramagnet,
\ie
    &H=\mathcal{U}_{CZ}H_0\,\mathcal{U}_{CZ}^\dagger\,,
\\
   &H_0=-\sum_{n}(X_{n}+\tilde{X}_n)+\text{h.c.}\,.\label{h_0}
\fe
The circuit $\mathcal{U}_{CZ}$ is composed of controlled-phase gates:
\begin{eqnarray}
\mathcal{U}_{CZ}&=&\prod_{n}[CZ_{n,n}]^{qb}[CZ^\dagger_{n,n+1}]^{q}\,.
\end{eqnarray}
Here, $CZ_{n,n}$ denotes the controlled-phase gate acting between the two $\mathbb{Z}_N$ qudits at site $n$ (i.e., the qudits with and without tilde), while $CZ_{n-1,n}$ denotes the controlled-phase gate acting between the qudit without tilde at site $n-1$ and the qudit with tilde at site $n$.
For instance, 
\ie
    CZ_{n,n}X_nCZ_{n,n}^\dagger&=X_n\tilde{Z}_n\,,\\
    CZ_{n,n+1}X_nCZ_{n,n+1}^\dagger&=X_n\tilde{Z}_{n+1}\,.
\fe
With this circuit, the ground state is simply given by
\ie
\ket{\text{GS}}=\mathcal{U}_{CZ}\left(\bigotimes_{n}(\ket{0}_n\otimes \ket{0}_n)\right)\,.
\fe
Here, the state $\ket{u}_n\otimes\ket{v}_n$ represents a simultaneous eigenstate of
$X_n$ and $\tilde{X}_n$,
\begin{eqnarray}
   X_n \ket{u}_n\otimes\ket{v}_n&=&\omega^u\ket{u}_n\otimes\ket{v}_n\,,\nonumber\\
\tilde{X}_n\ket{u}_n\otimes\ket{v}_n&=&\omega^v\ket{u}_n\otimes\ket{v}_n\,,
\end{eqnarray} 
where $\omega=e^{2\pi i/N}$.

The strong index of these SPT phases can be detected by placing the model on an open chain and investigating how the symmetry operators in Eq.~\eqref{symmetry} act near the boundaries. On an open chain of length $L$, using Eq.~\eqref{gs}, the symmetry operator when acting in the ground state subspace decomposes into boundary symmetry operators localized near the boundaries
\begin{eqnarray}
    \mathcal{U}_a\sim L_a\times R_a\,,\quad   \mathcal{U}_b\sim L_b\times R_b\,.
\end{eqnarray}
These boundary symmetry operators take the form 
\begin{eqnarray}
     L_a&=&\tilde{Z}_1^{\dagger abq}\,,\quad \qquad\quad\!\! L_b=\tilde X_1^b Z_1^{-pab^2}\,,\nonumber\\
     R_a&=&X^{a^{L}}_{L}\tilde{Z}_{L}^{q a^{L-1}}\,,\quad R_b=Z^{p b^L}\,.
\end{eqnarray}
and they form a projective algebra
\ie
L_aL_b
     &= \omega^{-\frac{b k N}{(ab-1,N)}}L_bL_a\,,
     \\
R_aR_b
     &= \omega^{\frac{b k N}{(ab-1,N)}}R_bR_a\,.   
\fe
Note that $b$ is invertible in $\mathbb{Z}_{(ab-1,N)}$ because it is manifestly coprime to $ab-1$, which is a multiple of $(ab-1,N)$. As a result, distinct $k\in \mathbb{Z}_{(1-ab,N)}$ give rise to distinct projective algebras on boundary and hence correspond to distinct strong SPT phases.

After discussing the construction of strong exponential SPT phases, we now discuss how to decorate them with weak indices. Recall that weak indices are specified by one-dimensional representations of $G_{\text{int}}$, which are labeled by $(s,t)\in\mathbb{Z}_N\times\mathbb{Z}_N$.
In our model, the weak indices are realized by modifying the strong SPT Hamiltonian in Eq.~\eqref{exspt} into
\ie
    H=-\sum_{n}\left[\omega^{t}Z_{n-1}^{p a}\tilde{X}_nZ_n^{\dagger p}+\omega^{s}\tilde{Z}_n^{ qb}X_n\tilde{Z}_{n+1}^{\dagger  q}\right]+\text{h.c.}\,.\label{weak}
\fe
The corresponding ground state is
\begin{eqnarray}
|\text{GS}\rangle =\mathcal{U}_{CZ}\left(\bigotimes_{n}(\ket{-s}_n\otimes \ket{-t}_n)\right)\,.\label{gsst}
\end{eqnarray}
Importantly, different $(s,t)$ do not represent distinct exponential SPT phases. This is because we can conjugate the Hamiltonian by a translation-invariant, exponential-symmetry-preserving, finite-depth quantum circuit 
\ie
\mathcal{U}=\prod_{n} Z_n^{ a} Z_{n+1}^{\dagger }=\prod_{n} Z_n^{(a-1)}
\fe
to shift $s$ by $(a-1)$. Hence, the  Hamiltonian with parameter $s$ describes the same SPT phase as that with $s+(a-1)$, leading to an equivalence relation $(s,t)\sim (s+(a-1),t)$. Similarly conjugated by 
\ie
\mathcal{U}=\prod_{n} \tilde Z_n^{ b} \tilde Z_{n+1}^{\dagger }=\prod_{n} \tilde Z_n^{(b-1)}
\fe 
shifts $t$ by $(b-1)$ and induces the equivalence relation $t\sim t+(b-1)$. Altogether, this yields the classification
$\mathbb Z_{(a-1,N)} \times \mathbb Z_{(b-1,N)}$,
which matches the result obtained from the cohomology analysis in Sec.~\ref{sec:CEP}.

To check that the Hamiltonian with different values of $s$ and $t$ indeed carries distinct weak indices, 
we insert an internal symmetry defect and examining how translation acts on the defect ground state. This is analogous to the standard diagnostic of 1+1D SPT phases via the insertion of symmetry defects. On an infinite chain, inserting a symmetry defect associated with the exponential symmetry $(g_1,g_2)$ between site $n_0-1$ and $n_0$ is realized by applying the symmetry transformation to all spins to the right of the defect. This modifies the Hamiltonian in Eq.~\eqref{weak} to
\ie
H_{(g_1,g_2)}=H
&+\left[\omega^{t}(\omega^{g_1 pa^{n_0}}\!-\!1)Z_{n_0-1}^{p a}\tilde{X}_{n_0}Z_{n_0}^{\dagger p}\!+\!\text{h.c.}\right]
\\
&+\left[\omega^{s}(\omega^{g_2qb^{n_0}}\!-\!1)\tilde{Z}_{n_0-1}^{ qb}X_{n_0-1}\tilde{Z}_{n_0}^{\dagger  q}\!+\!\text{h.c.}\right].
\fe
The corresponding ground state is
\ie
&|\text{GS}_{(g_1,g_2)}\rangle =
\\
&\mathcal{U}_{CZ}\, Z_{n_0}^{\dagger g_2qb^{n_0}}\!\tilde Z_{n_0+1}^{\dagger g_1pa^{n_0}}\!\left(\bigotimes_{n}(\ket{-s }_n\otimes \ket{-t}_n)\right).
\label{gsst}
\fe
In the presence of such a symmetry defect, the lattice translation operator has to be modified in order to preserve the location of the defect~\cite{10.21468/SciPostPhys.16.4.098,cheng2023lieb}. 
Denoting a lattice translation operator in the presence of $(g_1,g_2)$ symmetry defect between site $n_0-1$ and $n_0$ as $T^{(g_1,g_2)}_{\langle n_0,n_0+1\rangle}$, it should have the following form:
\ie
    T^{(g_1,g_2)}_{\langle n_0-1,n_0\rangle}={X}_{n_0-1}^{g_1a^{n_0}} \tilde{X}_{n_0-1}^{g_2b^{n_0}}T\,,\label{tr}
\fe
where $T$ is a lattice translation operator in the absence of the symmetry defect. 
We now examine how these modified translation operators act on the ground state of the weak SPT phases~\eqref{gsst}. One finds
\ie
T^{(g_1,g_2)}_{\langle n_0-1,n_0\rangle}|\text{GS}_{(g_1,g_2)}\rangle=\omega^{-g_1 s  a^{n_0}-g_2t  b^{n_0}}|\text{GS}_{(ag_1,bg_2)}\rangle\,,\label{pp}
\fe
which has the effect of mapping the  $(g_1,g_2)$ defect ground state to the $(ag_1,bg_2)$ defect ground state with a phase factor detecting the weak index. Shifting the defect by one lattice site changes the phase factor and effectively induces an equivalence relation $(s,t)\sim (a s, b t)$. 
Hence, distinct weak SPT phases are classified not by~$(s,t)$ itself, but by the equivalence classes under this transformation. This gives an alternative explanation to the classification $\mathbb Z_{(a-1,N)} \times \mathbb Z_{(b-1,N)}$ of the weak indices.

To summarize, we have demonstrated how our classification of the SPT phases works in the case of exponential symmetry and constructed a class of concrete lattice models, which exhausts all distinct exponential SPT phases. 
As a sanity check, when we set $a=b=1$, the exponential symmetry in Eq.~\eqref{symmetry} reduces to the standard uniform $\mathbb{Z}_N\times\mathbb{Z}_N$ symmetry with trivial modulation. Accordingly, 
the model in Eq.~\eqref{exspt} recovers the cluster states protected by the $\mathbb{Z}_N\times\mathbb{Z}_N$ global symmetry. Such a cluster state is labeled by $\mathbb{Z}_N\times \mathbb{Z}_N\times \mathbb{Z}_N$, 
where one $\mathbb{Z}_N$ factor corresponds to strong SPT phases classified by $H^2(G_{\text{int}},U(1))$, and the other two $\mathbb{Z}_N$ factors correspond to weak SPT phases classified by $H^1(G_{\text{int}},U(1))$.
This is consistent with Eq.~\eqref{80} where we set $a,b=1$.

\subsection{Dipolar SPT Phases}
We now consider the case of dipole symmetry. The internal symmetry is taken to be $G_{\text{int}}=\mathbb{Z}_N\times\mathbb{Z}_N$, with elements denoted by $g=(g_1,g_2)$. The lattice translation acts nontrivially on the internal symmetry as 
\begin{eqnarray}
    (g_1,g_2)\to  (g_1,g_2+g_1)\,.\label{dipsym}
\end{eqnarray}
For a finite periodic system of length $L$,  consistency requires $L=0\pmod N$. This type of modulated symmetry is referred to as \textit{dipole symmetry}, since the symmetry operators associated with $g_2$ is spatially uniform, while the ones associated with $g_1$ are dipole operators that depend linearly on the lattice site.
\subsubsection{Classification}
The strong index of SPT phases protected by the dipole symmetry is given by $H^2(G_{\mathrm{int}}, U(1))$, which happens to be invariant under the translation action in Eq.~\eqref{dipsym}.
Hence, valid modulated SPT is described by
\begin{eqnarray}
    \omega(g,h)=\exp\left(\frac{2\pi i}{N}kg_1h_2\right),\label{ppd}
\end{eqnarray}
where $g=(g_1,g_2)$, $h=(h_1,h_2)$, and $k\in\mathbb{Z}_N$. Indeed, $\omega(g,h)$ given in Eq.~\eqref{ppd} is invariant under the translation action in Eq.~\eqref{dipsym} up to coboundary
\ie
\frac{\omega(t(g),t(h))}{\omega(g,h)}=\exp\left(\frac{2\pi i}{N}kg_1h_1\right)=\frac{e^{i\theta(g+h)}}{e^{i\theta(g)}e^{i\theta(h)}}\,,
\fe
where $e^{i\theta(g)}$ is a function from $G$ to $U(1)$ of the form
\ie
e^{i\theta(g)}=
\begin{cases}
\exp\left(\dfrac{2\pi i}{2N} kg^2\right)\quad &\text{even }N\vspace{5pt}
    \\
\exp\left(\dfrac{2\pi i}{2N} kg(g-1)\right)\quad &\text{odd }N
\end{cases}\,.
\fe
This indicates that the strong dipolar SPT is characterized by $\mathbb{Z}_N$, in agreement with the classification previously obtained in \cite{lam2024classification}.\par
Weak SPT indices are given by one-dimensional representations of $G_{\text{int}}$, labeled by $(s,t)\in\mathbb{Z}_N\times\mathbb{Z}_N$
\begin{eqnarray}
    \chi_{s,t}(g)=\exp\left({\frac{2\pi i}{N}(sg_1+tg_2)}\right)\,.
\end{eqnarray}
However, 
two such representations may correspond to the same SPT phase if they are related by the action of translation. 
From Eq.~\eqref{dipsym}, the weak indices are subject to the following equivalence relation:
\begin{eqnarray}
    (s,t)\sim (s+t,t)\,\Longrightarrow\, (t,0)\sim (0,0)\,.
\end{eqnarray}
Quotienting by this relation gives the $\mathbb{Z}_N$ weak index.
\par
Overall, combining strong and weak indices, the dipolar SPT phases are classified by 
\begin{eqnarray}
   H^2(G_{\text{int}}\rtimes\mathbb{Z},U(1)) = \mathbb{Z}_N\times \mathbb{Z}_N\,. 
\end{eqnarray}
One can generalize the argument to the multi-flavor dipole symmetries. We set the internal symmetry to be
 \begin{eqnarray}
     G_{\text{int}}=(g^I_1,g^I_2)\in \prod_{I}\mathbb{Z}_{N_I}\times\prod_{I}\mathbb{Z}_{N_I}\,.\label{internalsym}
 \end{eqnarray}
Translation acts nontrivially on the internal symmetry as
 \begin{eqnarray}
     (g_1^I,g_2^I)\to(g_1^I,g_2^I+g_1^I)\,,\label{dipole algebra}
 \end{eqnarray}
which mixes the two components of the symmetry, realizing a nontrivial semidirect product structure. By the similar calculation as above, one can show that the SPT phases protected by this symmetry is classified by
  \begin{equation}
  \begin{aligned}
     &H^2(G_{\text{int}}\rtimes\mathbb{Z},U(1)) =
     \\
     &\prod_{I\leq J}\mathbb Z_{(N_{I},N_J)}\times \prod_{I< J}\mathbb Z_{(N_{I},N_J)}\times\prod_I\mathbb{Z}_{N_I}\,,
     \end{aligned}
 \end{equation}
 where the first two factors characterize strong dipolar SPT phases, while the last factor characterizes the weak ones.
The classification of strong dipolar SPT phases agrees with the one previously obtained in \cite{lam2024classification}.

\subsubsection{Lattice Models}\label{sec:dipole_lattice}
We now present a lattice-model construction for the dipolar SPT phases. Such a model with a nontrivial strong index was already studied in previous works~\cite{Han:2023fas, lam2024classification}. Here, we present a more complete model that incorporates both the strong and weak indices. 

We consider a spin chain where there is a $\mathbb{Z}_N$ spin degrees of freedom on each site. The $\mathbb{Z}_N$ dipole symmetry is realized as 
\begin{eqnarray}
    \mathcal{U}_C=\prod_{n}X_n,\quad \mathcal{U}_D=\prod_{n}X_n^n\,,\label{symd}
\end{eqnarray}
where $\mathcal{U}_C$ and $\mathcal{U}_D$ correspond to $g_2$ and $g_1$, respectively. Indeed, these symmetry operator transform under translation as in Eq.~\eqref{dipsym} 
\ie
T \,\mathcal{U}_C T^{-1}&= \mathcal{U}_C\,,
\quad
T \,\mathcal{U}_D T^{-1}&= \mathcal{U}_C \,\mathcal{U}_D\,.\label{eq:dipole_translation_algebar}
\fe

The dipolar SPT Hamiltonian is given by
\begin{eqnarray}
    H=-\sum_{n}\omega^{s}Z_{n-1}^q Z_n^{\dagger q}X_n Z_n^{\dagger q} Z_{n+1}^q+\text{h.c.}\,,\label{hamiltonian}
\end{eqnarray}
where $q\in\mathbb{Z}_N$ and $s\in\mathbb{Z}_N$ are the strong and the weak index, respectively, in the $\mathbb{Z}_N\times\mathbb{Z}_N$ classification we obtained before. The corresponding ground state reads
\begin{eqnarray}
|\text{GS}\rangle_{s} = \mathcal{U}_{CZ}\bigotimes_{n} |-s \rangle_n \,.
\end{eqnarray}
where $\mathcal{U}_{CZ}$ is a finite-depth quantum circuit of the form
\begin{eqnarray}
\mathcal{U}_{CZ}=\prod_{n}[CZ_{n,n+1}]^q [W^\dagger_{n}]^{q}\,,
\end{eqnarray}
with $CZ_{n,n+1}$ denoting the controlled-phase gate between sites $n$ and $n+1$, and $W_{n}$ denoting an operator acting on site $n$ defined by $W_n|s\rangle_{n'}=\omega^{s\delta_{n,n'}}|s\rangle_{n'}$.

Analogous to  the case of the exponential SPT phases, one finds that the strong index $q\in\mathbb{Z}_N$ enters into the boundary projective algebras of the symmetry operators~\eqref{symd}, while the weak index $s\in\mathbb{Z}_N$ determines the charge carried by the translation defects, consistent with the cohomology calculation. 

\subsubsection{Field Theory Perspective}
We  can formulate a continuum response theory for the dipolar SPT phases.  
 This construction provides a complementary field theoretic perspective on the classification derived from the previous analysis. 
Previous works have described dipole symmetries using
 higher derivative tensor gauge theories. Instead, we adopt an alternative description based on \textit{foliated gauge fields}~\cite{Ebisu:2023idd}, which provides a more direct connection with the group extension structure of the modulated symmetry. 
 \par
 To describe the response theory, we introduce background gauge fields for the dipole symmetry in Eq.~\eqref{internalsym} as
 ${a}^I$, $A^{I}$, where $a^I$ couples to the ordinary charge and $A^I$ couples to the dipole charge.
 The gauge fields are subject to the following gauge transformation~\cite{Ebisu:2023idd}:
\begin{eqnarray}
     a^I\to a^I+d\lambda^I+\Lambda^{I} e^x\,,\quad A^{I}\to A^{I}+d\Lambda^{I}\,,\label{gauge_transform}
 \end{eqnarray}
 where $e^x=dx$.
 In a continuum description, the dipole moment depends on the spatial separation between charges. To avoid ambiguities associated with this dependence, one may regard the theory as arising from a lattice regularization in which the lattice spacing sets the minimal dipole strength~\cite{PhysRevB.106.045112,Ebisu:2023idd}. In what follows, we take this unit length to be one. 
 It is convenient to introduce the following gauge field
 \begin{eqnarray}
 \tilde{a}^I:=a^I+xA^{I}\,.
 \end{eqnarray}
 Under the gauge transformation above, this field transforms as
 \begin{eqnarray}
     \tilde{a}^I\to \tilde{a}^I+d(\lambda^I+x\Lambda^{I})\,.
 \end{eqnarray}
We impose the following quantization conditions:
 \begin{eqnarray}
      \oint_{S_1}\tilde{a}^I=\frac{2\pi}{N_I}\mathbb{Z}\,,\quad \oint_{S_1} A^{I}=\frac{2\pi}{N_I}\mathbb{Z}\,.\label{loops}
 \end{eqnarray}
 These holonomies represent symmetry twists associated with the dipole symmetry.
 Since the symmetry is defined on a lattice, the coordinate $x$ should be regarded as a discrete lattice coordinate with unit spacing. Consequently, the combination $\tilde{a}^I=a+xA$ inherits the usual quantization condition associated with the~$\mathbb{Z}_{N_I}$ symmetry. 
 \par 
 Using these gauge fields, we can write the most general topological response action, corresponding to the second cohomology classes of the internal symmetries:
 \ie
     \mathcal L=&\sum_{I<J}\frac{k_{IJ}N_IN_J}{2\pi N_{IJ}}\tilde{a}^I\wedge \tilde{a}^J+\sum_{I, J}\frac{k^{\prime}_{IJ}N_IN_J}{2\pi N_{IJ}}\tilde{a}^{I}\wedge A^{J}\\
     &+\sum_{I<J}\frac{k^{\prime\prime}_{IJ}N_IN_J}{ 2\pi N_{IJ}}A^{I}\wedge A^{J}\,.\label{response}
 \fe
 Here, $k_{IJ}, k^\prime_{IJ}, k^{\prime\prime}_{IJ}\in\mathbb{Z}_{IJ}$, and $N_{IJ}=(N_I,N_J)$, (that is, gcd between $N_I$ and $N_J$). 
 However, the fields $\tilde{a}^I$ and $A^I$ are not independent because translation acts as $x\to x+1$, which induces
 \begin{equation}
     \tilde{a}^{I}\to \tilde{a}^{I}+ A^{I},\quad  A^{I}\to  A^{I}\,.\label{1-form}
 \end{equation}
 Applying this transformation to the response action in Eq.~\eqref{response}
 yields
 \begin{align}
     \begin{split}
         \delta\mathcal L=
         &\sum_{I< J}\frac{2 k_{IJ}N_IN_J}{2\pi N_{IJ}}\tilde{a}^{I}\wedge A^{J}
         -\sum_{I> J}\frac{2 k_{JI}N_IN_J}{2\pi N_{IJ}}\tilde{a}^{I}\wedge A^{J}\\
       &+\sum_{I<J}\frac{[k_{IJ}+( k^{\prime}_{IJ}-k^{\prime}_{JI})]N_IN_J}{ 2\pi N_{IJ}}A^{I}\wedge A^{J}\,.
     \end{split}
 \end{align}
 Requiring invariance of the action under translation therefore imposes the constraints
 \begin{equation}
     k_{IJ}=0,\quad k^{\prime}_{IJ}=k^{\prime}_{JI}\,.
 \end{equation}
 With these constraints, the response theory reduces to
 \ie
     \mathcal L=&\sum_{I\leq J}\frac{k^{\prime}_{IJ}N_IN_J}{2\pi N_{IJ}}\tilde{a}^{I}\wedge A^{J}+\sum_{I< J}\frac{k^{\prime}_{IJ}N_IN_J}{2\pi N_{IJ}}\tilde{a}^{J}\wedge A^{I}
     \\
     &+\sum_{I<J}\frac{k^{\prime\prime}_{IJ}N_IN_J}{ 2\pi N_{IJ}}A^{I}\wedge A^{J}\,,
 \fe
 which indicates that the strong index of the dipole SPT phases is labeled by $k^{\prime}_{IJ}\in \mathbb Z_{N_{IJ}}$ with $I\leq J$ and $k^{\prime\prime}_{IJ}\in \mathbb Z_{N_{IJ}}$ with $I< J$.
 \par
 Weak SPT phases correspond to stacking 0+1 SPT states along the spatial direction. A candidate response action is therefore described by
 \begin{eqnarray}
     \mathcal{L}_w=\sum_Ip_I\tilde{a}^I\wedge e^x+\sum_Ip^\prime_I  A^I\wedge e^x,
     \label{w}
 \end{eqnarray}
 where $p_I,p^\prime_I\in \mathbb{Z}_{N_I}$. Recalling that translation induces Eq.~\eqref{1-form}, we obtain
  $p_I=0$ by imposing translation invariance of the action. Hence, the weak index of the dipole SPT is labeled by $p^{\prime}_{I}\in\mathbb{Z}_{N_I}$. \par
 Combining the strong and weak indices, the dipolar SPT phases are classified by
 \begin{equation*}
     \prod_{I\leq J}\mathbb Z_{N_{IJ}}\times \prod_{I< J}\mathbb Z_{N_{IJ}}\times\prod_I\mathbb{Z}_{N_I},
 \end{equation*}
 which precisely matches the classification obtained from the lattice construction.

\section{Application}\label{sec:lsm}
\subsection{Lieb-Schultz-Mattis (LSM) Constraints}\label{sec:lsm_constraint}

In our MPS classification in Sec.~\ref{sec:MPS}, we assume that the local symmetry operators $U(a)$ form a linear representation of $G_{\text{int}}$. We now discuss what happens when the local symmetry operators instead form a projective representation characterized by the 2-cocycle $\nu(a,b)$
\ie
U(a)U(b)=\nu(a,b) U(ab)\,.
\fe
Recall that in a 1+1D system with both translation symmetry and an ordinary internal symmetry $G_{\text{int}}$ that does not mix with translation, such local projective representation leads to an LSM anomaly~\cite{Lieb1961,PhysRevB.69.104431,PhysRevLett.84.1535}, which implies that the system cannot admit a symmetric short-ranged entangled ground state and must instead either spontaneously break the symmetry or be gapless. 
However, as we will show this is not always the case for modulated symmetry, as previously pointed out in Ref.~\cite{pace2026lieb} using defect network. The condition we derive is the same one found in Ref.~\cite{pace2026lieb}.

When the local symmetry operators form a projective representation, following the same analysis that leads to Eq.~\eqref{eq:linearity}, we now arrive at
\ie\label{eq:projective}
\boxed{\frac{\omega(a,b)}{\omega(t(a),t(b))}=\nu(a,b)\frac{e^{i\theta_0(ab)}}{e^{i\theta_0(a)}e^{i\theta_0(b)}}\,. }
\fe
This means that $\nu (a,b)$ belongs to the same cohomology class as ${\omega(a,b)}/{\omega(t(a),t(b))}$. Given a specific $\nu(a,b)$, there may or may exist a cohomology class $\omega(a,b)$ that solves this condition. The solution does not exist if $\nu(a,b)\notin (T^*-1)H^2(G_{\text{int}},U(1))$. In this case, the local projective representation is intrinsically incompatible with the existence of modulated SPT phases, i.e., symmetric short-ranged entangled ground state, which implies an LSM anomaly of the modulated symmetry. On the other hand, if $\nu(a,b)\in (T^*-1)H^2(G_{\text{int}},U(1))$, there exists such an $\omega(a,b)$ that solves the condition. We can obtain the other solutions by multiplying $\omega(a,b)$ by a 2-cocycle $\Omega(a,b)$ whose cohomology class is invariant under translation. In this case, the classification of modulated SPT phases is identical to the one we obtained under the assumption of linear local representations. Although symmetric short-ranged entangled ground states are allowed, we can still deduce an SPT-LSM constraint on these ground state wave functions, analogous to the ones discovered in Ref.~\cite{Lu_2024,Jiang_2021,Pace_2025}, whenever $\nu(a,b)$ represents a nontrivial cohomology class. If $\nu(a,b)$ is nontrivial, any solution $\omega(a,b)$ to the condition must be a nontrivial 2-cocycle. Thus, the corresponding SPT phases must host nontrivial edge modes with projective symmetry action, and the ground state wave function must carry nontrivial entanglement due to the degeneracy in the entanglement spectrum. In particular, this means that a product state cannot be the ground state of a modulated SPT phase. As a result, there is no canonical notion of a trivial modulated SPT phase, and the set of the modulated SPT phases forms a torsor rather than a group.

\begin{figure}
    \centering
    \includegraphics[width=1\linewidth]{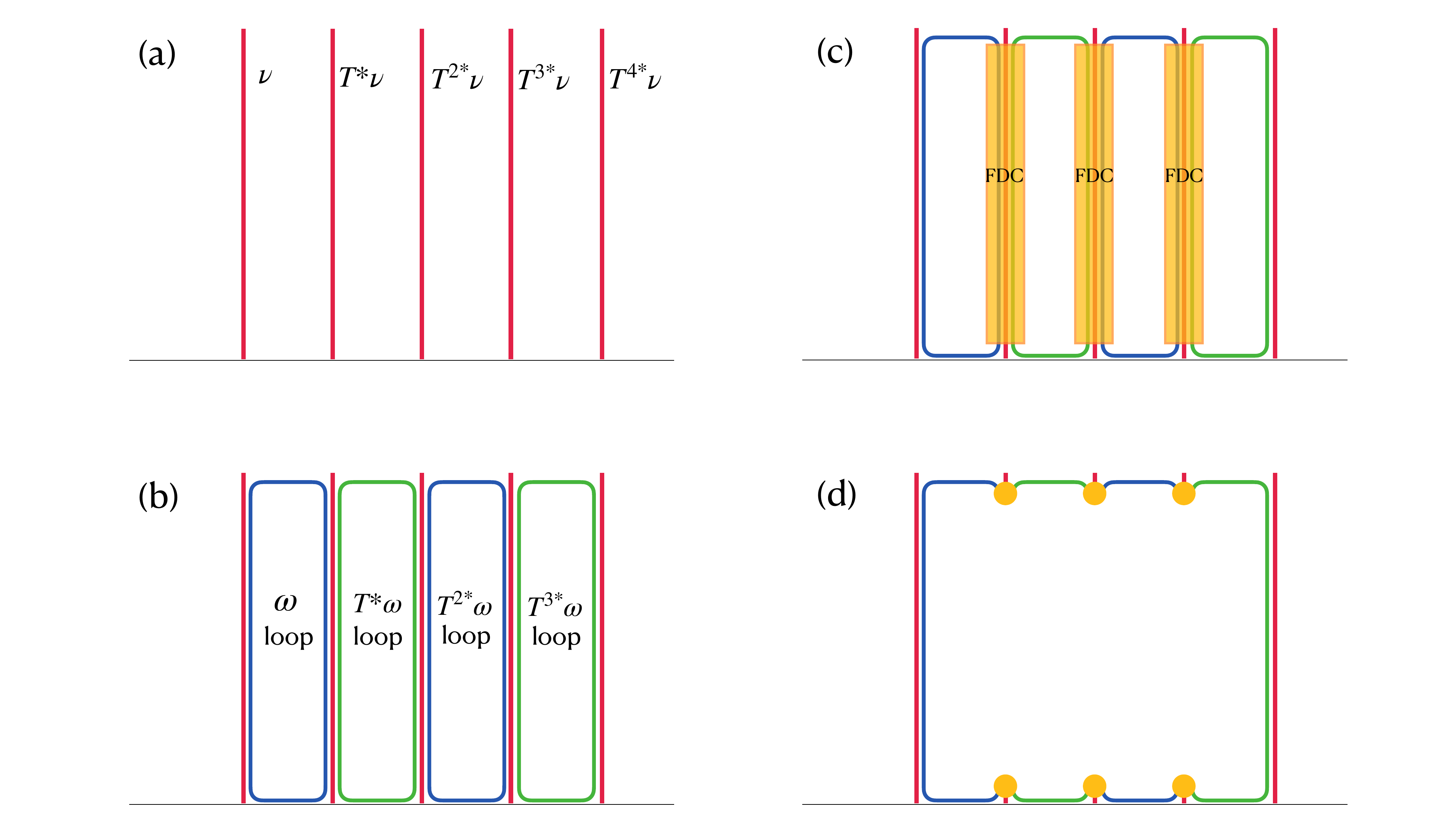}
    \caption{The trivialization process of 2+1D weak SPT phases and the necessary entanglement generated on the 1+1D boundary. (a) Illustration of a 2+1D weak SPT phase, which is a stack of 1+1D $G_{\text{int}}$ SPT phases, each supporting a projective representation at its edge.  Compatibility with the nontrivial action of translation on $G_{\text{int}}$ requires the stack to be modulated, i.e.~the 1+1D SPT phases transforms under translation.
    (b) We create loops of 1+1D $G_{\text{int}}$ SPT phases in the regions between the existing layers. This process can be achieved by a symmetric finite-depth quantum circuit. These SPT phases are also modulated so that translation symmetry is preserved.
    (c) As long as the 2-cocycle $\nu(a,b)$ satisfies Eq.~\eqref{eq:projective}, the three 1+1D SPT phases that assemble around the original $\nu(a,b)$ SPT phases (red) together form a total trivial SPT phase. We can therefore disentangle them by a symmetric finite-depth quantum circuit. (d) After removing the entanglement in the 2+1D bulk, all nontrivial entanglement remains on the boundary, which is exactly the 1+1D SPT-LSM state (bottom). This SPT-LSM state hosts a projective representation on its boundary characterized by $\omega(a,b)$ because the 1+1D $G_{\text{int}}$ SPT phases on the left and right boundaries of the 2+1D weak SPT phase are not completely trivialized.}
    \label{fig:SPT_LSM}
\end{figure}

We provide an alternative perspective on the LSM constraints we derive, based on anomaly inflow. In the modern viewpoint, the LSM constraint can be understood via anomaly inflow as arising from the boundary of a $2+1$D weak SPT phase, constructed by stacking 1+1D $G_{\mathrm{int}}$ SPT phases characterized by the 2-cocycle $\nu(a,b)$. This $2+1$D weak SPT is protected by the symmetry group $G = G_{\text{int}} \rtimes \mathbb{Z}$, where $\mathbb{Z}$ denotes the lattice translation symmetry.
If the $2+1$D weak SPT is nontrivial, its boundary carries the corresponding LSM anomaly, which forbids the existence of a symmetric short-range entangled ground state. Conversely, if it is a trivial SPT, there is no anomaly on the boundary and hence no obstruction to realizing a symmetric short-range entangled ground state.
For nontrivial $\nu(a,b)$, one can understand why a weak SPT phase obtained by stacking 1+1D SPT phases may nevertheless be trivialized by a mechanism illustrated in Fig.~\ref{fig:SPT_LSM}. This involves creating loops of SPT phases in the bulk and moving them toward the existing layers, so that their cohomology classes cancel, rendering the overall state trivial in $H^2(G, U(1))$. However, this process necessarily generates nontrivial entanglement on the boundary (see Fig.\ref{fig:SPT_LSM}). Physically, this can be understood in terms of a finite-depth quantum circuit that removes all bulk entanglement while leaving behind nontrivial entanglement at the boundary. 
Finally, we note that, via the crystalline equivalence principle, this mechanism corresponds to a trivialization map in the LHS spectral sequence. Related boundary states have also been referred to as anomalous SPT states in Ref.~\cite{QR21}.

To illustrate the LSM constraints with concrete examples, let us consider projective exponential symmetries and projective dipole symmetries.

For exponential symmetry, we consider the same $\mathbb{Z}_N\times\mathbb{Z}_N$ spin chain as in Sec.~\ref{sec:exp_lattice}, but with the following projective realization of the exponential symmetry:
\begin{eqnarray}
    \mathcal{U}_a=\prod_{n} \tilde Z_n^{ka^n}X_n^{a^n}\,,\quad     \mathcal{U}_b=\prod_{n} \tilde{X}_n^{b^n}\,,
\end{eqnarray}
where $k\in\mathbb{Z}_N$ determines the projectivity.
The algebra with the translation operator in Eq.~\eqref{eq:exp_algebra_translation} is preserved, so these operators indeed realize the exponential symmetry. The projectivity can be read off from the algebra of the local symmetry operators at site $n=0$, which gives
\ie
\nu(g,h)=\exp\left(\frac{2\pi i }{N} k g_1 h_2\right)\,.
\fe
The subgroup $(T^*-1)H^2(G_{\text{int}},U(1))$ is generated by
\ie
\omega(g,h)=\exp\left(\frac{2\pi i  }{N} (ab-1) g_1 h_2\right)\,.
\fe
Hence, if $k$ is not divisible by $\gcd(ab-1,N)$, the projective exponential symmetry carries an LSM anomaly that forbids a symmetric short-ranged entangled ground state. On the other hand, if $k$ is divisible by $\gcd(ab-1,N)$ and $k\neq 0$, the projective exponential symmetry allows the existence of a symmetric short-ranged entangled ground state, but it implies an SPT-LSM constraint that demands nontrivial entanglement in the ground state wavefunction and degenerate edge modes. This example has also been analyzed in Ref.~\cite{pace2026lieb} and our result is consistent with theirs.

For dipole symmetry, we consider the same $\mathbb{Z}_N$ spin chain as in Sec.~\ref{sec:dipole_lattice}, but with the following projective realization of the dipole symmetry:
\begin{eqnarray}
    \mathcal{U}_C=\prod_{n}X_n\,,\quad \mathcal{U}_D=\prod_{n}Z_n^kX_n^n\,,
\end{eqnarray}
which preserves the algebra with the translation operator in Eq.~\eqref{eq:dipole_translation_algebar},
where $k\in\mathbb{Z}_N$ determines the projectivity.
The projectivity can be read off from the algebra of the local symmetry operators at site $n=0$, which gives
\ie
\nu(g,h)=\exp\left(\frac{2\pi i }{N} k g_1 h_2\right)\,.
\fe
Since translation acts trivially on $H^2(G_{\text{int}},U(1))$, the subgroup $(T^*-1)H^2(G_{\text{int}},U(1))$ is trivial and hence $\nu(g,h)\notin (T^*-1)H^2(G_{\text{int}},U(1))$ when $k\neq 0$. This implies that the projective dipole symmetries carries an LSM anomaly whenever $k\neq 0$, which forbids a symmetric short-ranged entangled ground state.

\subsection{Anomalous Non-Invertible Kramers-Wannier Symmetries}

We now discuss another application of our classification of modulated SPT phases. In particular, we will use the classification to diagnose anomalies of non-invertible Kramers-Wannier symmetries in systems with modulated symmetries, and thereby establish an LSM-type constraints for these non-invertible symmetries. We focus on the case of $\mathbb{Z}_p$ exponential symmetry with prime $p$, although the analysis generalizes straightforwardly to other modulated symmetries.

In Ref.~\cite{Pace:2024tgk}, it was shown that the following Hamiltonian on a $\mathbb{Z}_p$ spin chain,
\ie\label{eq:ising}
H=-J\sum_n (Z_n Z_{n+1}^{-a^{-1}}+\text{h.c.})-K\sum_n (X_n+\text{h.c.})\,,
\fe
admits a non-invertible Kramers-Wannier reflection symmetry when $J=K$. This non-invertible symmetry is tied to the $\mathbb{Z}_p$ exponential symmetry preserved by this Hamiltonian, generated by
\ie
\mathcal{U}=\prod_n X_n^{a^n}\,.
\fe
This $\mathbb{Z}_p$ exponential symmetry forms a total symmetry group $G=\mathbb{Z}_p\rtimes \mathbb{Z}$ together with the lattice translation symmetry, where the translation generator $T$ acts on $g\in\mathbb{Z}_p$ as $T(g)=ag$. For the symmetry operator $\mathcal{U}$ to commute with the Hamiltonian $H$, we require $a^L= 1 \pmod p$ so that $\mathcal{U}$ is compatible with the periodic boundary condition, and we will consider periodic spin chains of such length $L$.

The non-invertible symmetry is a reflection symmetry that takes the form~\cite{Pace:2024tgk}
\ie
D=P\, R\, W \prod_{n=2}^{L} H_n\, CZ_n\,,
\fe
where $P$ is the projector onto the $\mathcal{U}=1$ subspace, $R$ is the reflection operator, $H_n$ is the Hadamard operator at site $n$, and $CZ_n$ is the controlled-phase operator acting on sites $n$ and $n-1$, obeying
\ie
&CZ_n \,X_n\, CZ_n^\dagger = Z_{n-1}^{a^{-1}} X_n\,,
\\
&CZ_n \,X_{n-1}\, CZ_n^\dagger = X_{n-1} Z_{n}^{a^{-1}}\,.
\fe
The operator $W$ acts on the sites $n=1$ and $n=L$ as
\ie
W \,X_1\, W^\dagger = Z_1^{-1} X_1 Z_1^{-1} Z_L^{a^{-1}}\,,\quad
W \,X_L\, W^\dagger = Z_1^{a^{-1}} X_L\,.
\fe
This operator $D$ is non-invertible due to the projector $P$. It implements a Kramers-Wannier transformation
\ie
D Z_n Z_{n+1}^{-a^{-1}} = X_{-n} D\,,\quad
D X_n =  Z_{-n}^{-1} Z_{-n+1}^{a^{-1}} D\,,
\fe
which leaves the Hamiltonian invariant when $J=K$.

With the complete classification of exponential SPT phases and their lattice realizations, we can show that this non-invertible symmetry is intrinsically incompatible with exponential SPT phases. According to our classification, the $\mathbb{Z}_p$ exponential SPT phases are classified by $\mathbb{Z}_{(a-1,p)}$. Since $H^2(\mathbb{Z}_p,U(1))=0$, there are no strong indices, and the classification arises entirely from weak indices, given by $H^1(\mathbb{Z}_p,U(1))/(1-T^*)=\mathbb{Z}_{(a-1,p)}$. These weak SPT phases are realized by
\ie
H_{\text{SPT}}=-\sum_n \bigl(\omega^k X_n+ \omega^{-k} X_n^\dagger\bigr)\,,
\fe
where $\omega= e^{2\pi i/p}$ and $k\in\mathbb{Z}_{(a-1,p)}$ parametrizes distinct SPT phases. It is clear that these Hamiltonians do not preserve the non-invertible Kramers-Wannier symmetry $D$, as the operator $D$ maps these SPT Hamiltonians to symmetry-breaking Hamiltonians,
\ie
&D H_{\text{SPT}} = H_{\text{SSB}} D\,,
\\
&H_{\text{SSB}}= -\sum_n \bigl(\omega^k Z_{n}^{-1} Z_{n+1}^{a^{-1}}+\omega^{-k} Z_{n} Z_{n+1}^{-a^{-1}}\bigr)\,.
\fe
We emphasize that although we demonstrated using the fixed-point Hamiltonian, the fact that $D$ maps an SPT phase to a symmetry-breaking phase does not depend on the details of the Hamiltonian. 
Therefore, there are no $\mathbb{Z}_p$ exponential SPT phases that preserve the non-invertible Kramers-Wannier symmetry. Consequently, this symmetry is anomalous, and any system with such a symmetry, e.g., the Hamiltonian in Eq.~\eqref{eq:ising} at $J=K$, must either be gapless or spontaneously break the non-invertible symmetry or the $\mathbb{Z}_p$ exponential symmetry.

\section{Conclusion}\label{sec:con}
We have studied SPT phases protected by modulated symmetries, where internal symmetry transform nontrivially under spatial symmetries, especially, lattice translation and reflection. Using the MPS formalism and the LHS spectral sequence, we provided a systematic and microscopic derivation of the classification of 1+1D modulated SPT phases and constructed an explicit map realizing the crystalline equivalence principle.

Within the MPS framework, we showed that the symmetry constraints on the MPS tensors naturally lead to a decomposition of the classification into strong and weak indices. The strong index can be detected either from the boundary symmetry algebra or from the  entanglement Hamiltonian, whereas the weak index does not affect these probes. Instead, it can be detected from the action of spatial symmetries in the presence of symmetry defects associated with the internal symmetry. When the spatial symmetry is lattice translation, the strong index corresponds to translation-invariant projective representations of the internal symmetry group, while the weak index arises from symmetry charges attached to unit cells and are defined modulo equivalence relations induced by translation. When the spatial symmetry is reflection, the strong index corresponds to reflection-reversed projective representations of the internal symmetry group, while the weak index corresponds to reflection-symmetric one-dimensional representations of the internal group modulo equivalence relations induced by reflection. The MPS analysis reveals a structure analogous to the LHS spectral sequence for the group cohomology of semidirect product groups, 
which enabled us to establish the cyrstalline equivalence principle.
\par
To demonstrate the physical realization of these phases, we constructed explicit lattice models realizing exponential and dipolar SPT phases and found that both the strong and weak indices coincide with the classification we derived. As applications of our classification, we analyzed the LSM anomalies and SPT-LSM constraints arising from modulated symmetries, and showed that a class of non-invertible Kramers-Wannier reflection symmetries is anomalous. 

Our results highlight the power of the MPS approach in characterizing modulated SPT phases. The framework developed here provides a useful starting point for exploring more general crystalline, subsystem, and higher-order multipole SPT phases.

\textit{Note added:}~During the preparation of this manuscript, we became aware of an upcoming related work~\cite{anakru2026matrix}, which may have overlap with some of our results.

\begin{acknowledgements}
SQN and HTL thank PiTP 2024:~Ultra-Quantum Matter for the hospitality, where this work was initiated.
HTL was supported by the U.S. Department
of Energy, Office of Science, Office of High Energy
Physics of U.S. Department of Energy under grant
Contract Number DE-SC0012567 (High Energy Theory research), by the Packard Foundation award for
Quantum Black Holes from Quantum Computation
and Holography and by the Simons Investigator Award
No. 926198.
HE is supported by JST CREST (Grant
No. JPMJCR24I3). 
SQN. is supported by a CRF from the Research Grants Council of the Hong Kong (No. C703722GF).
\end{acknowledgements}

\bibliography{MSPT}
\appendix 
\section{Lyndon–Hochschild–Serre Spectral Sequence} 
\label{sec:LHS}

In this appendix, we review the LHS spectral sequence following Ref.~\cite{QR21}, which relates the group cohomology $H^{p+q}(\tilde G,U(1)_s)$ for $\tilde G=G_{\text{int}}\rtimes \tilde G_{\text{sp}}$ to $H^{p}\big( \tilde G_{\text{sp}},H^q(G_{\text{int}},U(1))_s\big)$. For our interests of studying 1+1D modulated SPT phases, we specialize to the case of second group cohomology, where $p+q=2$, and take $G_{\text{int}}$ to be unitary.  We will follow the notation used in Sec.~\ref{sec:CEP}.

According to the LHS spectral sequence, the 2-cocycle of $\tilde G$ can be parameterized as 
\begin{align}
      F^{a_\mathbf{g},b_{\mathbf{h}}}=F^{{}^{\overline{\mathbf{gh}}}a,^{\overline{\mathbf{h}}}b}\times (F^{{}^{\overline{\mathbf{g}}}a,1_\mathbf{h}})^{s(\mathbf{g})} \times F^{1_\mathbf{g},1_\mathbf{h}} \,,\label{eq:ansatz}
\end{align}
and the sufficient conditions for $F^{a_\mathbf{g},b_\mathbf{h}}$ to be a 2-cocycle of $\tilde G$ are given as follows:
\begin{align}
    \frac{F^{b,c}\times F^{a,bc}}{F^{ab,c}\times F^{a,b}}&=1\,,
    \label{eq:LHS1_general}\\
       \frac{ (F^{{}^{\overline{\mathbf{g}}}a,{}^{\overline{\mathbf{g}}}b})^{s(\mathbf{g})}}{F^{a,b}}\frac{F^{b,1_\mathbf{g}}\times F^{a,1_\mathbf{g}}}{F^{ab,1_\mathbf{g}}}&=1\,,\quad\quad\ \, \label{eq:LHS_2_general}\\
\frac{ F^{a,1_\mathbf{hk}}}{F^{a,1_\mathbf{h}} (F^{{}^{\overline{\mathbf{h}}}a,1_\mathbf{k}})^{s(\mathbf{h})}}&=1\,,
   \label{eq:LHS3_general}\\
    \frac{ (F^{1_\mathbf{h},1_{\mathbf{k}})^{s(\mathbf{g})}} F^{1_\mathbf{g},1_{\mathbf{hk}}}}{F^{1_\mathbf{gh},1_{\mathbf{k}}} F^{1_\mathbf{g},1_{\mathbf{h}}}}&=1\,.
   \label{eq:LHS4_general}
\end{align}
These conditions exactly match those we derived from the MPS analysis in Eqs.~\eqref{eq:phy_gener_0}, \eqref{eq:phy_gener_1}, \eqref{eq:phy_gener_2}  and \eqref{eq:phy_gener_3} if we identify 
\ie
F^{a,b}=\omega(a,b)\,, \ \  F^{a,1_\mathbf{h}}=e^{-i\phi_\mathbf{h}({}^{\overline{\mathbf{h}}}a)}\,, \ \  F^{1_\mathbf{g},1_\mathbf{h}}=\omega(\mathbf{g},\mathbf{h})\,.
\fe
Furthermore, substituting these identifications back into Eq.~\eqref{eq:ansatz} leads to the same parametrization as in Eq.~\eqref{eq:parametrization_2_cocycle} if we identify $\omega(a_\mathbf{g},b_\mathbf{h})=F^{a_\mathbf{g},b_\mathbf{h}}$.

The LHS spectral sequence has a physical interpretation in terms of domain wall decoration \cite{QR21}. In particular, $F^{a,1_\mathbf{g}}$ characterizes the decoration of 0+1D $G_{\text{int}}$ SPT on the fluctuating $\mathbf{g}$ domain wall. This is more explicitly when we choose $F^{a,b}=1$, then on the $\mathbf{g}$ domain wall of the $G$ SPT, there is a one-dimensional representation of $G_{\text{int}}$ defined by $\mu_\mathbf{g}(a):=F^{a,1_\mathbf{g}}$.  The condition Eq.~\eqref{eq:LHS3_general} becomes the compatible condition on the one-dimensional representation under the domain wall fusion.

\end{document}